\documentclass[fleqn,usenatbib]{mnras}

\usepackage{newtxtext,newtxmath}
\usepackage[T1]{fontenc}
\usepackage{longtable}
\usepackage{ulem}
\usepackage{xcolor}

\DeclareRobustCommand{\VAN}[3]{#2}
\let\VANthebibliography\thebibliography
\def\thebibliography{\DeclareRobustCommand{\VAN}[3]{##3}\VANthebibliography}

\usepackage{graphicx}	% Including figure files
\usepackage{amsmath}	% Advanced maths commands
\title[A revised moonlit sky brightness model]{A revised simplified scattering model for the moonlit sky brightness profile based on photometry at SAAO}

\author[H. Winkler]{
Hartmut Winkler,$^{1}$\thanks{E-mail: hwinkler@uj.ac.za}
\\
$^{1}$Dept. Physics, University of Johannesburg, Kingsway, 2006 Auckland Park, Johannesburg, South Africa\\
}

\date{Accepted XXX. Received YYY; in original form ZZZ}

\pubyear{2021}

\begin{document}
\label{firstpage}
\pagerange{\pageref{firstpage}--\pageref{lastpage}}
\maketitle

\begin{abstract}
This paper presents multi-filter measurements of the night sky brightness at the South African Astronomical Observatory (SAAO) in Sutherland in the presence of a bright moon. The observations cover a wide range of sky directions, lunar phases and lunar positions. A revised simplified scattering model is developed for estimating the sky brightness due to moonlight that more accurately reflects the atmospheric extinction of the lunar beam compared to models frequently applied in astronomical studies. Contributions to night sky brightness due to sources other than moonlight are quantified and subtracted from the total sky background radiation to determine the spectral intensity and angular distribution of scattered moonlight. The atmospheric scattering phase function is then derived by comparing the sky brightening to the strength of the incoming lunar beam, estimated using a novel approach. The phase function is shown to be an excellent match to the combined theoretical Rayleigh and Mie scattering functions, the latter with a Henyey--Greenstein form instead of the exponential angular relationship often used in previous studies. Where deviations between measured and model sky brightness are evident in some bands these are explained by contributions from multiple scattering or airglow, and are quantified accordingly. The model constitutes an effective tool to predict sky brightness at SAAO in optical photometric bands, especially with a bright moon present. The methodology can also be readily be adapted for use at other astronomical sites. The paper furthermore presents $UBV(RI)_c$ and Str{\"o}mgren photometry for 49 stars, most with no prior such data.
\end{abstract}

\begin{keywords}
atmospheric effects -- site testing -- techniques: photometric -- Moon -- stars: general
\end{keywords}

%%%%%%%%%%%%%%%%%%%%%%%%%%%%%%%%%%%%%%%%%%%%%%%%%%

%%%%%%%%%%%%%%%%% BODY OF PAPER %%%%%%%%%%%%%%%%%%

\section{Introduction}

Night sky brightness determines the degree of background noise in all types of optical astronomical measurements. Its accurate estimation is thus critical in determining the instrumental configuration and integration times required to achieve scientifically useful astronomical observations. This is particularly true for programmes at large modern facilities requiring extensive observational pre-planning and using semi-automated telescope scheduling.

This has by necessity led to a significant growth in night sky characterisation studies at many major astronomical facilities, including Calar Alto \citep{Leinert1995, Sanchez2007}, ESO-La Silla \citep{Mattila1996}, La Palma \citep{Benn1998}, Cerro Tololo \citep{Krisciunas2007}, ESO-Paranal \citep{Patat2003, Patat2008, Noll2012} and Xinglong \citep{Yao2013}.

By far the biggest factor determining the night sky brightness (outside twilight times) is the Moon. It affects roughly half of all astronomical observations. Its degree of impact depends both on accurately determinable parameters such as the lunar phase and position in the sky, as well as far less certain factors associated with local conditions, particularly the nature, concentration and distribution of the small particles suspended in the atmosphere referred to as aerosols.

Despite the substantial and frequently dominant contribution of moonlight to astronomical observations, especially around the period of the full moon, surprisingly little work has been carried out in quantifying this component, and characterising it in terms of wavelength and other parameters. It is remarkable that the study by \citet{Krisciunas1991}, which was based on just 33 ad-hoc photometric measurements, all in just the $V$-band, was the seminal work in this field for over 20 years. A subsequent series of papers on the night sky at Cerro Paranal \citep{Noll2012,Jones2013,Jones2019} are the first major developments in this field of study since then. While \citet{Noll2012} merely applied the Krisciunas \& Schaefer methodology with modified parameters, \citet{Jones2013} introduced a sophisticated moonlight model based on modern spectral atmospheric radiative transfer codes that incorporated multiple scattering.

Despite these developments, simplified moonlight models remain useful in practice. This paper will demonstrate how a straightforward correction to the Krisciunas \& Schaefer formulation leads to adequate moonlight sky profiles for photometric studies, particularly for shorter wavelength pass bands. The research presented here partly draws on prior studies of solar irradiance, particularly daytime sky brightness models based on the single scattering of sunlight \citep[see e.g.][]{Kocifaj2009}. The paper evaluates night sky brightness measurements at the South African Astronomical Observatory (SAAO) main observational site 15 km west of the town of Sutherland. Site studies at SAAO-Sutherland have largely focused on astronomical seeing and extinction \citep{Catala2013,Kilkenny1995}. The site's night sky brightness has however not been investigated in a comprehensive, systematic manner to date. Night sky illumination there is almost entirely due to natural light sources, and this will continue to be the case for the foreseeable future \citep{Sefako2015}.

\section{The night sky}

\subsection{Summary of night sky contributors}
\label{sec:summary}

The brightness of the night sky is determined by a combination of diverse components. The total intensity $I$ of light measured by an observer at ground level emanating from a specific small patch in the sky is then 
\begin{equation}
I = \sum I_i = I_\text{T} + I_\text{L} + I_\text{S} + I_\text{Z} + I_\text{A} + I_\text{P} + I_\text{O}
\end{equation}
where the individual components $I_\text{T}$, $I_\text{L}$, $I_\text{S}$, etc. are described in the sub-chapters that follow.

\subsubsection{Twilight}

Scattered sunlight brightens the sky for considerable periods before sunrise and after sunset. Twilight periods are traditionally classified according to the solar zenith angle $Z$, and are defined as: i) civil twilight - $90\degr < Z < 96\degr$ , i.e. the Sun is $0\degr$-$6\degr$ below the horizon, ii) nautical twilight - $96\degr < Z < 102\degr$ (Sun $6\degr$-$12\degr$ below the horizon), and iii) astronomical twilight - $102\degr < Z < 108\degr$ (Sun $12\degr$-$18\degr$ below the horizon).

During civil twilight the scattered sunlight component $I_\text{T}$ dominates all other components, and even during nautical twilight it remains prominent and variable on short timescales. These periods are therefore not suitable for the study of the fainter night sky components. During astronomical twilight, the overall rate of sky brightness change decreases, as the contribution from weaker but steadier contributors to the sky luminance become more important. In the presence of a bright moon, the scattered sunlight component becomes relatively small during this period, but cannot be ignored entirely.

\subsubsection{Scattered moonlight}

When the moon is near or above the horizon then the scattered lunar light intensity component $I_\text{L}$ is usually dominant outside twilight times, particularly near full moon. As this is the prime subject of this paper, scattered moonlight will be discussed fully in the later sections.

\subsubsection{Integrated starlight}

In this study care is taken to exclude faint foreground stars from the sky brightness measurements, so this component $I_\text{S}$ refers to starlight from distant, unresolved sources, and notably grows in strength towards lower galactic latitudes, where it constitutes the Milky Way. Away from the galactic plane, this component is roughly inversely proportional to the sine of the galactic latitude. Its spectrum there is effectively an integrated spectrum of starlight in our Galaxy. Its strength as a function of wavelength and galactic latitude $b$ has been determined by \citet{Mattila1980}, and the results of that study will be applied here. Closer to the galactic plane, interstellar extinction becomes significant.

\subsubsection{Zodiacal light}

This component corresponds to sunlight scattered off minute granules in solar system interplanetary space. Its spectrum broadly resembles the solar spectrum. It grows in strength with decreasing ecliptic latitude and as the ecliptic longitude approaches the position of the Sun. Intensity due to zodiacal light $I_\text{Z}$ as a function of sky coordinates has been parameterised by several authors, including \citet{Leinert1998}, whose determination forms the basis of the analysis that follows.

\subsubsection{Airglow}

Airglow is the result of upper atmosphere transitions in atoms and molecules such as atomic oxygen, $\text{O}_2$, OH and FeO \citep{Hart2019, Unterguggenberger2017}. It is notably variable, with a dependence on solar activity, seasonal trends, short term periodicities (the latter typically of the order of 5-10 minutes, with amplitudes of 2-5 per cent) as well as frequent night-long downward drifts \citep{Leinert1998, Patat2003, Patat2008}. Many long-term investigations however highlight how different line sets exhibit different behaviour, longer variational timescales with larger amplitudes, as well as frequent deviations from average trends \citep[e.g.][]{Marsh2006, Hart2019}. This variability makes airglow almost impossible to predict accurately without an actual measurement thereof. However, the spectral lines that collectively constitute airglow are found almost exclusively in the red part of the night sky spectrum. For dark skies they have a major effect on the $I_c$-band sky brightness, and a smaller influence on the $V$ and $R_c$ magnitudes.

When neglecting Earth curvature effects, the dependence of the airglow intensity with zenith angle $\zeta$ may be approximated by the expression $I_\text{A}(\zeta) = I_\text{A}(\zeta = 0\degr)\sec{\zeta}$ \citep{Noll2012}.

\subsubsection{Light pollution}

The scattering of light due to artificial sources can dramatically increase sky brightness, usually in and around urban centres. The basis of the theory used to quantify the intensity due to light pollution $I_\text{P}$ is captured in the work of \citet{Krisciunas1991} and previous papers cited there. More recent developments in the study of light pollution include theoretical computational modelling \citep[e.g.][]{Kerola2006} and field surveys \citep[e.g.][]{Falchi2011}.

\subsubsection{Other components}

Other contributors $I_\text{O}$ were assumed to have negligible impact on the results of this study compared to moonshine. One of these is auroral light, which only becomes significant at high latitudes. Others include emission from other galaxies and extragalactic background light.

\subsection{Light scattered in the atmosphere from an external point source}
\label{sec:scatter}

In the atmosphere a beam of light of flux $F$ and intensity $I$ is diluted by scattering and absorption. If a beam's initial intensity is $I_\text{i}$ upon entering a uniform section of length $s$, then the intensity at exit $I_\text{f}$ is determined by the relation d$I = -I \kappa$ d$s \Rightarrow I_\text{f} = I_\text{i} \exp (-\kappa s)$, with an analogous expression applying for $F$ as well. $\kappa$ is referred to as the volume extinction coefficient, which describes both the concentration and scattering efficiency of the intervening particles, and is generally wavelength dependent.

It is convenient to measure the effective path length in terms of the relative airmass, which is approximated here as the secant of the zenith angle, i.e. neglecting Earth curvature effects. The flat atmosphere simplification is justified in this work in view of the fact that stars could not be observed below $30\degr$ above the horizon due to the instrumental limitations. Due to this approximation, the treatment of a beam's traverse of the generally non-uniform atmosphere now becomes mathematically equivalent to moving through a uniform layer with thickness measured in units of the vertical airmass above the observing site.

\begin{figure}
	\includegraphics[width=\columnwidth]{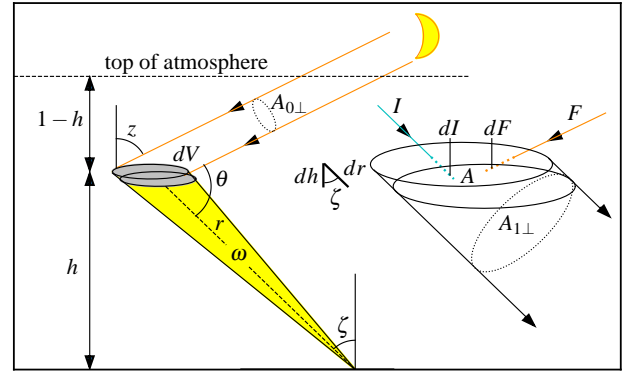}
    \caption{Schematic diagram illustrating the incident and scattered lunar beam and
     associated parameters.}
    \label{fig:parameters}
\end{figure}

The model that will be applied here is partly based on parameters illustrated in Fig.~\ref{fig:parameters}. In this construct we imagine the atmosphere as a semi-transparent, uniform, constant density air layer between a ground level height of $h = 0$ and the top of the atmosphere at $h = 1$ vertical airmass. When adopting relative airmass ($= \sec\zeta$) as the length measure, the volume scattering coefficient becomes the vertical optical depth $\tau$, which hereafter will just be referred to as optical depth. This parameter is widely utilised in atmospheric science, and is related to the extinction coefficient $k$ more commonly used in observational optical astronomy by the identity $\tau = 0.921k$ \citep[e.g.][]{Formenti2002}, where $k$ is measured in magnitudes. Then
\begin{equation}
    \text{d}I = -I \tau \sec\zeta \text{d}h \Rightarrow I(h) = I(h_\text{i}) \exp{(- \tau \sec\zeta \Delta h)} \, .
	\label{eq:ext}
\end{equation}

In what follows the intensity of the scattered moonlight from a patch of the sky is initially determined under the simplifying assumption that the contribution to sky brightness due to photons scattered into the line of sight as a result of second or subsequent scattering events can be neglected. This implies that multiple scattering of a photon is rare. That is the case in situations where the optical depth is small, which is indeed typical of conditions experienced at a good astronomical site such as SAAO where the data was obtained for this paper. Despite this assumption, the model presented here does incorporate the further attenuation of photons that have been scattered once due to a second scattering or absorption event.

The revised simplified scattering model developed in what follows (hereafter RSS model) therefore considers reflected skylight as primarily the product of an incident direct beam from a single bright source (in this case, the Moon) that has been scattered just once. Both the incident direct beam and the scattered beam are however subject to atmospheric attenuation along the entire path used to traverse the atmosphere. That was also the approach adopted by \citet{Krisciunas1991}, but there is one critical change in the model formulation developed here that corrects for an underestimation of the direct lunar beam intensity reaching the scattering point by \citeauthor{Krisciunas1991} and other subsequent investigators \citep{Noll2012, Yao2013, Chakraborty2005}. The modification is explained in full later on. Other differences introduced here are that the calculations in this paper utilise a normalised scattering function, a different expression for the aerosol scattering angular dependence and the incorporation of an empirical estimate of the multiple scattering contribution.

The RSS model presented here considers a beam of direct (unscattered) moonlight incoming at a zenith angle $z$ being scattered inside a volume element d$V$ with an angle $\theta$ towards the observer. This volume element is initially envisaged to be a thin cylinder of base area $A$ and height d$h$, i.e. d$V = A$d$h$, set in the horizontal plane. This alignment is chosen for mathematical convenience, and it can be shown that the measured sky brightness is independent of the shape and orientation of the element d$V$ \citep[see e.g.][]{Horvath2014}. The height $h$ of element d$V$ above the ground, and $r$, the distance from element d$V$ to the observer, are measured in units of airmass.

The probability that a photon will be scattered into a particular direction is determined by the scattering function, referred to as $f(\theta)$ by \citet{Krisciunas1991}. Specifically note that the functions $f(\theta)$ determined by \citet{Krisciunas1991} and \citet{Noll2012} are not normalised, and rather include scaling parameters $C_\text{R}$ and $C_\text{M}$ required to incorporate conversions between non-SI units for sky brightness and lunar flux.

When the scattering function is normalised, this is then referred to as the phase function $p$. It describes the probability $p(\theta,\phi) $d$\Omega$ that a scattered photon is propelled into the direction defined by the solid angle d$\Omega$ \citep[see e.g.][]{Kocifaj2009}. The phase function has units of sr$^{-1}$, and its integral over all solid angles equates to 1. While this will not be applied here, a $4\piup$ factor is sometimes included in this definition, in which case that integral would be $4\piup$.

The notation that will be employed in what follows uses the subscript "0" for radiation associated with the incident lunar beam, "1" for radiation of photons that have experienced one single scattering event only, and "2" where more than one scattering event has taken place. The subscript "s" signifies that the quantity refers to scattering only, and excludes absorption losses.

$A_{0 \perp}$ and $A_{1 \perp}$ are the cross-sectional areas of d$V$ when viewed from the direction of the incoming beam and cone $\omega$ respectively. We immediately note that $A = A_{0\perp} \sec{z} = A_{1\perp} \sec{\zeta}$. If a linear beam of incoming radiation with flux $F_0$ is incident on the thin volume element d$V$, then the radiative power lost by the beam due to scattering inside d$V$ will be d$P_0 = A_{0 \perp}$ d$F_\text{0s}$, while the power of the radiation scattered in this process into solid angle $d\Omega$ will be d$P_1 = p$ d$P_0$ d$\Omega$. The corresponding scattered flux will be d$F_1 =$ d$P_1 / A_{1 \perp}$. The solid angle associated with d$V$ as viewed by the observer, $\omega = A_{1\perp}/r^2$, can be set to the angular size of the sky element used to determine the sky brightness. In the data analysis that follows, $\omega$ will be chosen to equate to 1 arcsec$^2$.

If $p$ is furthermore independent of the azimuthal angular coordinate $\phi$, then the parameter relationships introduced in the previous paragraph yield that
\begin{equation}
    p(\theta) = \frac{\text{d}P_1}{\text{d}P_0 \text{d}\Omega} = \frac{A_{1 \perp}}{A_{0 \perp}} \frac{\text{d}I_1}{\text{d}F_\text{0s}} \, ,
	\label{eq:phasef}
\end{equation}
\noindent since, by the definition of irradiance, $I =$ d$F /$ d$\Omega$. The argument above is independent of the shape of d$V$, and it follows that we may choose this to take the form of a flat surface of area $A$ and thickness d$h$ as illustrated in Fig.~\ref{fig:parameters}.

When reaching the volume d$V$, the lunar beam has already been attenuated along the effective path $(1-h)\sec{z}$ it has traversed inside the atmosphere up to this point. Thus the flux of the lunar beam $F_0$ just before entering d$V$ is
\begin{equation}
    F_0 = F_{\text{L}}^* \exp{(- \tau (1-h) \sec{z})} \, ,
	\label{eq:lunarflux}
\end{equation}
where $F_{\text{L}}^*$ is the lunar beam flux at the top of the atmosphere. The scattering losses to the lunar beam in traversing d$V$ will then be d$F_\text{0s} = F_0 \tau_\text{s} \sec{z}$ d$h$.

Equation (\ref{eq:lunarflux}) marks a decisive deviation of the RSS model from the formulation used by \citet{Krisciunas1991} -- from now on referred to as the KS model -- and later studies based on it \citep{Noll2012, Yao2013}. The KS model approach did not include the $(1-h)$ factor, and effectively assumes that the lunar beam traverses the entire atmosphere from top to ground prior to scattering (instead of only the upper part of the atmosphere down to element d$V$). According to the KS model, the lunar flux before scattering is therefore less than it really is, and this leads to an underestimation of the sky brightness. This is illustrated clearly later in Section~\ref{sec:KScomp}.

From equations (\ref{eq:phasef}) and (\ref{eq:lunarflux}), the intensity of radiation scattered towards the observer inside d$V$ is therefore
\begin{multline}
    \text{d}I_1 = p(\theta) \frac{A \cos{z}}{A \cos\zeta} F_{\text{L}}^* \tau_\text{s} \sec{z} \exp{(- \tau (1-h) \sec{z})} \text{d}h \\
    = p(\theta) F_{\text{L}}^* \tau_\text{s} \sec\zeta \exp{(- \tau (1-h) \sec{z})} \text{d}h\, .
\end{multline}

Not all of this will however reach ground level due to attenuation along the path $r$ in the cone defined by solid angle $\omega$ highlighted in Fig.~\ref{fig:parameters}. The total intensity of scattered moonlight detected at ground level was earlier labeled $I_\text{L}$. This is constituted by moonlight undergoing just a single scattering event, $I_\text{L1}$, and moonlight scattered more than once, $I_\text{L2}$. So $I_\text{L} = I_\text{L1} + I_\text{L2}$.

d$I_{L1}$, the intensity of lunar light scattered inside d$V$ reaching ground level bounded by $\omega$ and being subjected to further attenuation inside the cone, is
\begin{multline}
    \label{eq:dI1}
\text{d}I_\text{L1} = \text{d}I_1 \times \text{e}^{-{\tau}r} \\
 = p(\theta) F_{\text{L}}^* \sec\zeta \tau_\text{s} \text{e}^{-\tau((1-h)\sec{z} + h\sec{\zeta})} \text{d}h \, .
\end{multline}
Hence, integrating over all possible volume elements d$V$ that make up cone $\omega$ from top to bottom of the atmosphere, we get
\begin{multline}
    \label{eq:IL1}
    I_\text{L1} = p(\theta) F_{\text{L}}^* \tau_\text{s} \sec{\zeta} \times \text{e}^{- \tau \sec{z}} \int_{0}^{1} \text{e}^{-\tau(\sec{\zeta} - \sec{z})h}\text{d}h \\
    \Rightarrow I_\text{L1} = p(\theta) F_{\text{L}}^* \frac{\tau_\text{s}}{\tau} \sec{\zeta} \frac{\text{e}^{-\tau \sec{\zeta}} - \text{e}^{-\tau \sec{z}}}{\sec{z} - \sec{\zeta}} \, .
\end{multline}

This result is already well established in analogous determinations of the scattering of sunlight in the atmosphere \citep[e.g][]{Dubovik2000, Kocifaj2009, Kittler2012, Horvath2014}. 

So the scattering phase function (in logarithmic terms) is expressed by
\begin{equation}
    \log{p(\theta)} = \log{I_\text{L}} - \log{F_{\text{L}}^*} - \log{\Gamma} - \delta \, ,
\label{eq:logphase}
\end{equation}
where:
\begin{equation*}
    \Gamma = \sec{\zeta} \frac{\text{e}^{-\tau\sec{\zeta}} - \text{e}^{-\tau\sec{z}}}{\sec{z}-\sec{\zeta}} \, .
\end{equation*}
is referred to as the gradation function \citep{Kocifaj2009}. Note that $\Gamma$ reduces to $\tau \sec{\zeta} e^{-\tau \sec\zeta}$ if $\zeta = z$. Furthermore,
\begin{equation*}
    \delta = \log \left[ \frac{\tau_\text{s}}{\tau} \eta \right] \, ,
\end{equation*}
where $\tau_\text{s} / \tau$ represents the average albedo of atmospheric particles and $\eta = I_\text{L} / I_\text{L1}$ constitutes the ratio of total scattered to single scattered radiation. $\eta$ is mostly dependent on the Rayleigh optical depth \citep[e.g.][]{Leinert1998}, meaning that the term $\delta$ is effectively a shift in the $\log p$ expression that incorporates scattering parameters, but is assumed to show no significant zenith angle dependence.

As the sky patch surface magnitude was determined in units of magnitude per arcsec$^2$ while $p$ is measured in sr$^{-1}$, a dimensional conversion for these units will be incorporated in the calculations which follow that in the logarithmic form used in Equation~\ref{eq:logphase} amounts to $-10.629$.

\subsection{Empirical scattered moonlight models}

Previous work considered both Rayleigh and Mie scattering mechanisms. In the case of the former, which is associated with small scattering particles and is thus appropriate for molecules as found in air, \citet{Krisciunas1991} deduced a scattering function $f_\text{R}(\theta) = C_\text{R} (1.06+\cos^2\theta)$ with $C_\text{R} = 2.27 \times 10^5$. Note that, apart from the scaling factor, this relation is consistent with the normalised theoretical Rayleigh scattering function given by \citet{Bucholtz1995}, which translates to the scattering phase function
\begin{equation}
    p_\text{R}(\theta) = \frac{1}{4\piup} \frac{3(1 - \chi)}{4(1 + 2\chi)} \left[ \frac{1+3\chi}{1-\chi} + \cos^2\theta \right] \, .
\end{equation}
$(1 + 3\chi)/(1 - \chi) = 1.06$ corresponds to a value of $\chi = 0.0148$, in full agreement with the $0.01384 < \chi < 0.01557$ tabled for the wavelength range covered in the \citeauthor{Bucholtz1995} study. Tests with different $\chi$ values shown later do not suggest any reason to deviate from $\chi = 0.0148$, which is therefore adopted here again.

The normalised Mie scattering function, which is appropriate for aerosols with sizes of the order of a photon's wavelength, is often approximated by a Henyey--Greenstein function
\begin{equation}
    \label{eq:HG}
    p_\text{M}(\theta) = \frac{1}{4\piup} \frac{1-g^2}{(1+g^2-2g\cos{\theta})^{3/2}} \, ,
\end{equation}
where $g$ is referred to the asymmetry factor \citep{Henyey1941}, which depends on aerosol type, but is typically of the order of 0.5 for Earth's atmosphere.

\citet{Krisciunas1991} however chose the (once again not normalised) expression
\begin{equation}
    f_\text{M}(\theta) = C_\text{M} 10^{-\theta/40} \, ,
\end{equation}
with $\theta$ given in degrees and with $C_\text{M} = 10^{6.15}$, as best describing their measurements obtained at Mauna Kea. In their study of the moonlit night sky over Cerro Paranal, \citet{Noll2012} found that the mathemetical form of the expressions for $f_\text{R}$ and $f_\text{M}$ also matched their data. Their results however required revised constants of $C_\text{R} = 10^{5.70}$ and $C_\text{M} = 10^{7.15}$.

The RSS model introduced here instead characterises Mie scattering utilising the more common Henyey--Greenstein phase function (equation~(\ref{eq:HG})), as this produces a better match to the data presented in this paper.

When we combine the Rayleigh and Mie phase functions, their relative fraction will be determined by the ratio of the scattering optical depth for the Rayleigh ($\tau_\text{R}$) and Mie ($\tau_\text{M}$) processes respectively to the total $\tau_\text{s} = \tau_\text{R} + \tau_\text{M}$ value. The combined normalised phase function is then
\begin{equation}
\label{eq:phase}
    p(\theta) = \left[ \frac{\tau_\text{R}}{\tau_\text{s}} p_\text{R} (\theta) + \frac{\tau_\text{M}}{\tau_\text{s}} p_\text{M} (\theta) \right] \, .
\end{equation}

The Rayleigh scattering optical depth can be estimated with the expression
\begin{equation}
    \label{eq:tau_R}
    \tau_{\lambda,\text{R}} = \frac{P}{P_0} (1.229 \times 10^{10}) \lambda^{-4.05} \, ,
\end{equation}
\noindent where $P$ is the observing site ground level atmospheric pressure (in mb), $P_0 = 1013.5$ mb is standard atmospheric pressure at sea level and the wavelength $\lambda$ is in nanometers \citep{Dutton1994}.

\section{Observation and data analysis}

\subsection{Photometry}

The experimental strategy adopted was based on the available instrumentation, which required calibration through regular photometric observations of bright standard stars. In order to maximise the number of sky brightness measurements these were performed in suitable sky patches in the immediate vicinity of standard stars and other stars with accurately known photometric magnitudes. The sky measurements then also serve for the purpose of sky subtraction, thus speeding up the calibration star observing sequence.

Observations were carried out on the 0.5 m telescope at the South African Astronomical Observatory in Sutherland during two observing runs in October/November 2012 and March 2013. The telescope was equipped with a modular photometer and a Hamamatsu R943-02 gallium arsenide photomultiplier. The filters used included both Johnson--Cousins $UBV(RI)_c$ as well as Str{\"o}mgren $uvby$ sets. All star and sky measurements were made through a 45 arcsec diameter circular aperture. The observing sequence adopted entailed measuring the star for 10\,s in $B$, $V$, $R_c$, $I_c$ and 20\,s in $U$, then measuring the sky background three times in each filter ($2{\times}10$\,s plus $1{\times}5$\,s in $B$, $V$, $R_c$, $I_c$ and $2{\times}20$\,s plus $1{\times}10$\,s in $U$), ending with another measurement in each filter of the star (using the same integration times as in the first star measurement). In the case of the Str{\"o}mgren photometry, an analogous observing sequence was used, with integration times now 10\,s in $b$, $y$, 15\,s in $v$ and 20\,s in $u$ for each of the two star measurements, and $2{\times}10$\,s plus $1{\times}5$\,s in $v$, $b$, $y$ and $2{\times}20$\,s plus $1{\times}10$\,s in $u$.

Low-level spurious noise of instrumental, atmospheric or cosmic origin could significantly impact on the faint sky measurements. The advantage of having three sky measurements in each filter is that any dubious reading is readily identified as an outlier. Less than 2 per cent of the readings were rejected for this reason.

The instrument's dark counts were determined by taking a multitude of readings with all shutters closed and lights switched off in the dome. In view of the cooling of the photomultiplier tube, these counts were found to be significantly lower than the counts obtained from even the darkest sky patches. The dark count rate determined in this manner was always subtracted from the sky patch readings.

Standard stars from the E- and F-regions were observed to calibrate the photometric measurements, using $UBV(RI)_c$ magnitudes from the list of \citet{Menzies1989} and $uvby$ magnitudes from \citet{Kilkenny1992}. These covered a wide range of airmass values, enabling an accurate determination of the photometric extinction coefficients $k_\lambda$. The $k_\lambda$ proved to be consistent with SAAO-Sutherland's default extinction values for all nights, which were therefore adopted for all calculations.

In view of the blue nature of the reflected skylight preference was given to blue stars with a magnitude brighter than $V = 8$\,mag, as for these the stellar flux is still clearly distinguishable above the sky flux, even when the star is less than $10\degr$ from the Moon. As blue stars are frequently variable, B-stars shown to be adequately stable in previous studies \citep{Menzies1990, Menzies1991, Winkler1989, Winkler1990, Winkler1997} were preferred. It was further required that these stars had no bright companions 20-30 arcsec away and that uncrowded patches of sky of at least 1 arcmin in diameter could be identified in the immediate neighbourhood of these stars.

In order to convert the measured magnitudes to fluxes, the standard definition $m = -2.5 \log(F_{\lambda}/F_{\lambda,0})$ was applied utilising zero-magnitude fluxes $F_{\lambda,\text{cal}}$ and effective wavelengths $\lambda_{\text{eff}}$ for each filter from the calibrations of \citet{Bessell2012} for $UBV(RI)_c$ and \citet{Gray1998} for $uvby$. These are summarised in Table~\ref{tab:filters} for convenience.

\begin{table}
	\centering
	\caption{Effective wavelength (in nm) and magnitude-flux conversion factors (in units of $\times 10^{-9}$\,erg\,s$^{-1}$cm$^{-2}$\AA$^{-1}$) for the Johnson--Cousins filters \citep{Bessell2012} and Str{\"o}mgren filters \citep{Gray1998}.}
	\label{tab:filters}
	\begin{tabular}{lrrrrr}
	\hline
		& $U$ & $B$ & $V$ & $R_c$ & $I_c$ \\
		$\lambda_\text{eff}$ & 367.3 & 436.8 & 545.5 & 642.6 & 793.9 \\
		$F_{\lambda,\text{cal}}$ & 4.176 & 6.386 & 3.685 & 2.206 & 1.178 \\
	\hline
		& $u$ & $v$ & $b$ & $y$ & \\
		$\lambda_\text{eff}$ & 349.1 & 411.1 & 466.2 & 545.6 & \\
		$F_{\lambda,\text{cal}}$ & 11.72 & 8.66 & 5.89 & 3.73 & \\
	\hline
	\end{tabular}
\end{table}

Table~\ref{tab:listshort} (and \ref{tab:starlist}) lists the observed stars together with their coordinates, spectral types and where available a reference to previous photometry. If a star was a photometric standard then their E- or F-region star identification number is listed instead. The offset in RA and Dec of the associated sky patch is presented in columns 6 and 7 (RA(sky patch) = RA(star) + $\Delta$RA, Dec(sky patch) = Dec(star) + $\Delta$Dec).

\begin{table}
	\centering
	\caption{List of target stars with associated sky patches.The complete list is available as Table~\ref{tab:starlist}.}
	\label{tab:listshort}
\begin{tabular}{lrlllrr}
\hline
RA2000 & Dec2000 & Name & MK & Ref & $\Delta$RA & $\Delta$Dec \\
hh\,mm\,ss & $\degr\,\arcmin\,\arcsec$ & & type & & s & \arcmin \\
\hline
00 07 37 & $-$86 02 20 & HD 385    & B9IV     & [W97] &  0 & +1.5 \\
00 13 56 & $-$17 32 43 & HD 955    & B4V      &       & +5 & +1.3 \\
00 15 57 &   +04 15 04 & HD 1160   & A0V      &       & +5 & $-$1.1 \\
00 48 48 &   +18 18 50 & HD 4670   & B9       &       & $-$6 & $-$1.1 \\
00 56 49 & $-$25 21 48 & HD 5524   & A2/3V    &       & +6 & +1.5 \\
... & ... & ... & ... & ... & ... & ... \\
\hline

\multicolumn{7}{l}{Notes: [W97] - \citet{Winkler1997}} \end{tabular}
\end{table}

\subsection{Lunar brightness}
\label{sec:lunbright}

The determination of the lunar brightness at a particular point in time is complex, and requires not only accurate Sun and Moon positions and distances relative to the observer's exact location, but also knowledge of the lunar reflective properties, which vary widely over the surface of the Moon and are in addition a function of the solar beam incident angle on the Moon's surface and the subsequent scattering angle off the lunar ground. \citet{Jones2013} have developed a methodical though for some purposes laborious process for estimating this lunar component that includes determination of predicted double and multiple scattering fractions.

This paper follows a similar approach, though with different procedures in places. Here, as a starting point, the solar magnitude (at 1 AU) was taken to be $V = -26.71$ mag \citep{Pecaut2013}. From this, the solar magnitude in the other Johnson--Cousins bands were determined from the solar colours given by \citet{Ramirez2012}:
$(B - V) = 0.653$, $(U - B) = 0.166$, $(V - R_c) = 0.352$, $(V - I_c) = 0.702$. Using the conversion equation $V = y + 0.08 (b - y)$ \citep{Kilkenny1992}, the corresponding Str{\"o}mgren magnitude is estimated to be $y = -26.74$ mag. From this the solar magnitudes in the other Str{\"o}mgren bands were obtained using the colours given by \citet{Melendez2010}: $(b - y) = 0.4105$, $(v - b) = 0.6227$, $(u - v) = 0.9546$. The solar magnitudes and corresponding fluxes actually measured at the lunar surface are then easily determined by applying the conversion factors in Table 1, the Sun-Moon distance in AU and the $F \propto 1/(\text{distance})^2$ relation.

\begin{figure}
	\includegraphics[width=\columnwidth]{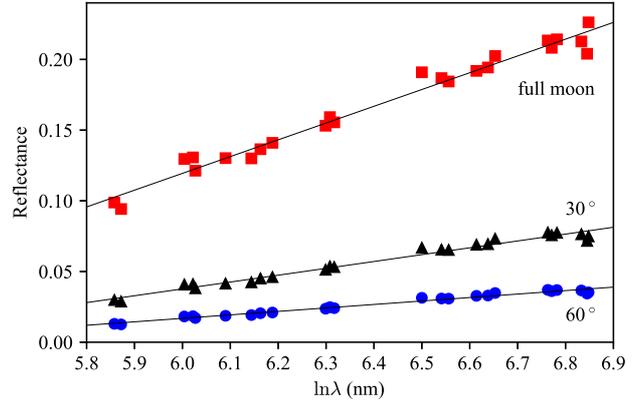}
    \caption{Examples of a disk-equivalent lunar reflectance vs. $\ln \lambda$ graph based on the \citet{Kieffer2005} calibration, displaying the linear nature of this graph. The squares, triangles and circles correspond to phase angles $0\degr$, $30\degr$ and $60\degr$ from full moon, and the selenographic longitude and latitude of the observer were here set to $0\degr$.}
    \label{fig:lunaralbedo}
\end{figure}

As in \citet{Jones2013}, the work of \citet{Kieffer2005} was used to determine the lunar albedo in the observer's direction. Kieffer and Stone presented complex empirical relationships determining the Moon's reflective properties for a set of specific wavelengths in the optical range. The formulas defining these relationships not only depend on lunar phase, but also the selenographic coordinates of both the Sun and observer. \citet{Kieffer2005} provide expressions and sets of coefficients for determining what they refer to as the disk-equivalent lunar reflectance at specific wavelengths. This quantity was calculated at the time of each sky patch observation. It was noticed that when the albedo determined by these relationships is plotted against the logarithm of the wavelength, these graphs are very well described by a linear function (see Fig.~\ref{fig:lunaralbedo} for three examples). This was found to be true for all lunar phases and alignments encountered in the present study. In contrast to \citet{Jones2013}, these linear fits were then used to determine the lunar albedo and hence the lunar magnitude at the effective wavelength corresponding to each of the filters.

Having thus calculated the fraction of sunlight reflected off the Moon's surface at each filter's effective wavelength, the intensity of moonlight at the top of the atmosphere is then determined by applying the flux-distance relationship to the Moon-observer distance at the time of observation.

\subsection{Determining the sky light components}
\label{sec:components}

\subsubsection{The lunar fraction}

If all the relevant observing site-specific parameters are known, the sky brightness due to scattered moonlight can be determined by calculating $I_\text{L1}$ with equation (\ref{eq:IL1}) and multiplying this quantity by $\eta$. As these parameters have not all been established yet, the quantity $I_\text{L}$ needs to be obtained from the measured sky surface magnitude $\mu$ and estimates of the other sky brightness components. As it is assumed in this study that all contributors other than $I_\text{L}$, $I_\text{Z}$, $I_\text{S}$ and $I_\text{A}$ can be neglected, we have that
\begin{equation}
\label{eq:lunarI}
   I_\text{L} = F_{\text{cal}} \times 10^{-0.4 \mu} - (I_\text{Z} + I_\text{S} + I_\text{A}) \, .
\end{equation}

\subsubsection{Zodiacal light}

The contribution of the zodiacal light was determined using the calibration and wavelength dependence relationships given by \citet{Leinert1998}. The ecliptic longitude (relative to the solar ecliptic longitude at the time of observation) and ecliptic latitude of each target were computed using standard astrometric formulae. The zodiacal light contribution at 500\,nm was then estimated through linear interpolation on a relative ecliptic longitude vs. ecliptic latitude table of the logarithm of the values given in table 16 of \citet{Leinert1998} (converted to units of W\,m$^{-2}$\,arcsec$^{-2}$). The (top of the atmosphere) zodiacal light intensity $I_{\text{Z}}^*$ for each filter's effective wavelengths was thereafter calculated by applying a filter-specific correction factor determined through linear interpolation between the wavelength dependence relationships for specific ecliptic longitudes presented in equation 22 of \citet{Leinert1998}. For sky patches within $30\degr$ of the ecliptic poles $I_{\text{Z}}^*$ was also adjusted to account for the Earth's position relative to the interplanetary dust plane using the calibration given in Equation 17 of \citet{Masana2021}. The ground level intensity is then obtained with $I_\text{Z} = I_{\text{Z}}^* \exp(-\tau_\text{eff} \sec\zeta)$, where $\tau_\text{eff}$ is the optical depth adjusted for scattering into the line of sight. $\tau_\text{eff}$ is related to the actual optical depth by $\tau_\text{eff} = f_\text{eff} \tau$, where $f_\text{eff}$ is determined from an empirical relationship described by \citet{Noll2012}.

\subsubsection{Integrated starlight}

The integrated starlight component was determined using table III of \citet{Mattila1980}, which estimates the (top of the atmosphere) intensity $I_{\text{S}}^*$ due to this component for specific wavelengths and galactic latitudes $b$ assuming a galactic extinction to distance ratio of 2.0\,mag\,kpc$^{-1}$ for the $V$ band. For each filter, the integrated starlight intensity for $\csc|b|$ = 1.00, 1.25, 1.67, 2.50, 5.00 and 6.00 was determined by linearly interpolating the filter effective wavelength between the entries (converted to units of W\,m$^{-2}$\,arcsec$^{-2}$) for the next lowest and next highest wavelengths in the table. The galactic latitude dependence for that filter was then obtained through a quadratic fit to the associated $I_{\text{S}}^*$ vs. $\csc|b|$ plot: 
\begin{equation}
    I_{\text{S}}^* = (a_\text{S0} + a_\text{S1} \csc|b| + a_\text{S2} \csc^2|b|) \times 10^{-18} \, \text{W\,m$^{-2}$\,arcsec$^{-2}$} \, ,
\end{equation}
where $a_\text{S0}$, $a_\text{S1}$ and $a_\text{S2}$ are listed in Table~\ref{tab:zsaparms}. The quadratic functions obtained in this manner match the data of \citeauthor{Mattila1980} very well, but would not be appropriate for low galactic latitudes, where the starlight component is much larger and very irregular. For this reason sky patches at $|b| < 8\degr$ were excluded in later calculations. Here too $I_\text{S} = I_{\text{S}}^* \exp(-\tau \sec\zeta)$. Note that again there is some scattering of this component into the beam \citep[see e.g.][]{Noll2012}, mostly near the galactic poles. The effect is however relatively minor, especially when compared to other uncertainties (e.g. airglow fluctuations), and will therefore be ignored here.

\begin{table*}
	\centering
	\caption{Integrated skylight and airglow fitting parameters adopted for the $UBV(RI)_c$ and $uvby$ filters.}
	\label{tab:zsaparms}
	\begin{tabular}{llcccccccccc}
	\hline
		Filter & & $U$ & $B$ & $V$ & $R_c$ & $I_c$ & & $u$ & $v$ & $b$ & $y$ \\
		\hline
		starlight & $a_\text{S0}$ & $-0.2186$ & $-0.4181$ & $-0.4540$ & $-0.4416$ & $-0.4200$ & & $-0.2167$ & $-0.3968$ & $-0.4820$ & $-0.4540$ \\
		 & $a_\text{S1}$ & +1.1047 & +1.9550 & +2.1618 & +2.0117 & +1.7603 & & +0.9957 & +1.8213 & +2.2676 & +2.1618 \\
		 & $a_\text{S2}$ & $-0.06223$ & $-0.10682$ & $-0.10473$ & $-0.08550$ & $-0.05928$ & & $-0.05524$ & $-0.09872$ & $-0.12086$ & $-0.10473$ \\
		airglow & $a_\text{A}$ & 0.62 & 0.26 & 1.69 & 3.44 & 10.55 & & 0.67 & 0.48 & 0.08 & 2.86 \\
		\hline
	\end{tabular}
\end{table*}

\subsubsection{Airglow}

The highly variable airglow component originates at very high altitudes that can be treated as top of the atmosphere, and was determined using
\begin{equation}
    I_{\text{A}}^* = a_\text{A} \times 10^{-18} \, \text{W\,m$^{-2}$\,arcsec$^{-2}$} \, ,
\end{equation}
where the values of $a_\text{A}$ are listed in Table~\ref{tab:zsaparms}. These values were estimated by generating the zenith spectrum for the airglow continuum plus upper air emission lines with the ESO {\sc{SKYCALC}} Sky Model Calculator \citep{Noll2012,Jones2013} for the Cerro Paranal site (set for all year and all night average and a water vapour column density of 2.5 mm). The contribution of this airglow template to each filter was thereafter established by multiplying the spectrum with the filter transmission functions (taken from \citet{Bessell2012} for the Johnson--Cousins bands and from \citet{Bessell2011} for the Str{\"o}mgren filters). The airglow measured at ground level is once more also affected by airglow photons scattered into the line of sight, and so $I_\text{A} = I_{\text{A}}^* \exp(-\tau_\text{A,eff} \sec\zeta)$. Here too $\tau_\text{A,eff} = f_\text{A,eff} \tau$. $f_\text{A,eff}$ is a function of airmass, and has been empirically determined by \citet{Noll2012}.

\subsection{Calculating the scattering phase function}

Table ~\ref{tab:filtparms} displays the optical depth values appropriate for SAAO-Sutherland. As the standard photometric extinction coefficients $k_\lambda$ proved suitable for all observations described here, the total optical depth simply equates to $0.921 k_\lambda$ \citep{Formenti2002}.

The Rayleigh scattering optical depth $\tau_\text{R}$ only depends on the observing site's atmospheric pressure. This may for the purposes of this analysis be taken to be 825 mb, which is typical of the values recorded by the meteorological station close to the SAAO telescopes. The resulting Rayleigh optical depths at the wavelengths adopted for the various filters, calculated with equation (\ref{eq:tau_R}), are then given in Table~\ref{tab:filtparms}.

\begin{table*}
	\centering
	\caption{Optical depth and $\delta$ (with standard deviation -- SD) for the $UBV(RI)_c$ and $uvby$ filters.}
	\label{tab:filtparms}
	\begin{tabular}{llcccccccccc}
	\hline
		Filter & & $U$ & $B$ & $V$ & $R_c$ & $I_c$ & & $u$ & $v$ & $b$ & $y$ \\
		\hline
		total & $\tau_{\lambda}$ & 0.497 & 0.249 & 0.138 & 0.092 & 0.064 & & 0.562 & 0.295 & 0.184 & 0.129 \\
		Rayleigh & $\tau_{\lambda,\text{R}}$ & 0.409 & 0.203 & 0.082 & 0.042 & 0.018 & & 0.503 & 0.259 & 0.156 & 0.082 \\
		aerosol & $\tau_{\lambda,\text{M}}$ & 0.045 & 0.032 & 0.020 & 0.015 & 0.010 & & 0.050 & 0.036 & 0.028 & 0.020 \\
		gases & $\tau_{\lambda,\text{G}}$ & 0.043 & 0.014 & 0.036 & 0.035 & 0.036 & & 0.009 & 0.000 & 0.000 & 0.027 \\
		offset & $\delta_{\lambda}$ & 0.426 & 0.270 & 0.078 & $-0.027$ & $-0.198$ & & 0.479 & 0.338 & 0.249 & 0.137 \\
		offset SD & $\sigma(\delta_{\lambda})$ & 0.056 & 0.048 & 0.069 & 0.103 & 0.196 & & 0.056 & 0.044 & 0.045 & 0.057 \\
		\hline
	\end{tabular}
\end{table*}

The Mie scattering optical depth $\tau_\text{M}$ depends on both the aerosol concentration and particle properties. The analysis here uses the values given in Table~\ref{tab:filtparms}, which were found to be representative of conditions encountered on most days at SAAO \citep{Formenti2002}.

The remaining component $\tau_\text{G}$ making up the total optical depth is due to absorption from atmospheric gases such as ozone and water vapour. This too is listed in Table~\ref{tab:filtparms}, and was derived from the other values in the Table using $\tau_\text{G} = \tau - \tau_\text{R} - \tau_\text{M}$.

With the proportion of Rayleigh and Mie scattering events determined, the normalised phase function (equation (\ref{eq:phase})) can now be established. This can be matched to the plot of $p(\theta)$ obtained from the photometric measurements through equation (\ref{eq:logphase}), initially treating the unknown offset $\delta$ as zero. In all instances the plot profile matches the form of the theoretical phase function, but with a vertical offset between the two. This offset is then removed through adjusting $\delta$ for each filter until the median offset between individual points and the theoretical curve became zero.

\section{Results}

\subsection{Stellar photometry}

While this study primarily sought to measure sky brightness, the observations produced as a by-product a series of serendipitous photometric measurements for a range of stars that are not photometric standards. In many cases these have no previous observations in these bands. The data are compiled in Table~\ref{tab:JCshort} (Table~\ref{tab:JClong}) and Table~\ref{tab:stromshort} (Table~\ref{tab:stromlong}).

\begin{table}
	\centering
	\caption{$UBV(RI)_c$ magnitudes of observed stars. The second column lists the mid-point in time of each observation $D$, in fractional days, relative to the heliocentric Julian day (HJD) 2456000.0, i.e. $D = \mbox{HJD} - 2456000$. The complete version is available as Table~\ref{tab:JClong}.}
	\label{tab:JCshort}
\begin{tabular}{lccrrrr}
\hline
Name & $D$ & $V$ & $B-V$ & $U-B$ & $V-R_c$ & $V-I_c$ \\
\hline
HD 385    & 225.384 & 7.40 & $-$0.01 & $-$0.16 &    0.00 &    0.01 \\
HD 955    & 225.498 & 7.36 & $-$0.16 & $-$0.64 & $-$0.08 & $-$0.16 \\
          & 232.428 & 7.39 & $-$0.14 & $-$0.64 & $-$0.08 & $-$0.16 \\
HD 1160   & 226.342 & 7.10 &    0.04 &    0.04 &    0.00 &    0.01 \\
HD 4670   & 227.369 & 7.91 &    0.00 & $-$0.16 & $-$0.01 &    0.00 \\
... & ... & ... & ... & ... & ... & ... \\
\hline
\end{tabular}
\end{table}

\begin{table}
	\centering
	\caption{Results of the Str{\"o}mgren photometry of the observed stars. As in Table~\ref{tab:JCshort}, the second column lists $D = \mbox{HJD} - 2456000$. In line with common practice, the $u$ and $v$ magnitudes are not given explicitly, but rather in terms of the indexes $m_1 = (v-b) - (b-y)$ measuring metallicity and $c_1 = (u-v) - (v-b)$ quantifying the Balmer jump. The complete version is available as Table~\ref{tab:stromlong}.}
	\label{tab:stromshort}
\begin{tabular}{lccrrr}
\hline
Name & $D$ & $y$ & $b-y$ & $m_1$ & $c_1$ \\
\hline
HD 955    & 232.433 & 7.43 & $-$0.07 & 0.10 & 0.34 \\
HD 1160   & 226.348 & 7.11 &    0.00 & 0.19 & 0.99 \\
HD 4670   & 227.376 & 7.94 &    0.02 & 0.11 & 0.92 \\
HD 5524   & 227.390 & 7.22 &    0.05 & 0.19 & 1.05 \\
HD 6815   & 232.332 & 7.30 & $-$0.03 & 0.12 & 0.84 \\
... & ... & ... & ... & ... & ... \\
\hline
\end{tabular}
\end{table}

\subsection{Sky patch surface magnitudes}

Table~\ref{tab:UBVRIshort} (for $UBV(RI)_c$; full version: Table~\ref{tab:UBVRIlong}) and Table~\ref{tab:uvbyshort} (for $uvby$; full version: Table~\ref{tab:uvbylong}) list the positional parameters and surface magnitudes of the sky patches determined at the time of measurement. In column 1, the sky patches have been labeled according to the HD catalogue number of the associated star (i.e. the sky patch adjacent to the star HD 385 has been named "SP 385", etc.). Column 2 lists the day and time of observation, while the lunar phase, as seen at the location and time of measurement, is given in column 3 as the fraction of the lunar disk illuminated by the Sun as viewed by the observer, ranging from 0 (new moon) to 1 (full moon). Columns 5 and 6 list the lunar zenith angle $z$ and azimuth $\gamma$ (measured from due north in an initially easterly direction, i.e. anticlockwise for an observer facing the zenith at the observing site). Columns 6 and 7 give the sky patch zenith angle $\zeta$ and azimuth $\alpha$ (measured as for the lunar azimuth). The angular separation $\theta$ between the Moon and the sky patch, as seen by the observer, is given in column 8. The remainder of the table is made up of the sky patch surface magnitude (per arcsec$^2$) and colours.

\begin{table*}
	\centering
	\caption{$UBV(RI)_c$ surface magnitudes of the observed sky patches. Columns 3-8 respectively list the lunar phase, zenith angle and azimuth, sky patch zenith angle and azimuth, and scattering angle at the time of observation. The complete version is given in the Appendix (Table~\ref{tab:UBVRIlong}).}
	\label{tab:UBVRIshort}
\begin{tabular}{lccrrrrrcrrrr}
\hline
Sky patch & $D$ & lunar & $z$ & $\gamma$ & $\zeta$ & $\alpha$ & $\theta$ & $\mu(V)$ & $B-V$ & $U-B$ & $V-R_c$ & $V-I_c$ \\
label &  & phase & ($\degr$) & ($\degr$) & ($\degr$) & ($\degr$) & ($\degr$) & (mag/$\sq\arcsec$) & (mag) & (mag) & (mag) & (mag) \\
\hline
SP 385    & 225.3840 & 0.7745 & 45.2 & 300.5 & 53.6 & 180.9 &  82.15 & 19.87 & 0.16 & $-$0.38 & 0.18 & 0.99 \\
SP 955    & 225.4980 & 0.7823 & 77.7 & 273.4 & 47.4 & 275.7 &  30.36 & 19.48 & 0.31 &    0.01 & 0.20 & 0.85 \\
          & 232.4279 & 0.9594 & 58.0 &  27.4 & 32.3 & 289.8 &  67.19 & 19.47 & 0.05 & $-$0.38 & 0.02 & 0.47 \\
SP 1160   & 226.3422 & 0.8527 & 34.4 & 338.6 & 37.0 &   9.2 &  17.94 & 18.77 & 0.25 & $-$0.20 & 0.14 & 0.55 \\
SP 4670   & 227.3685 & 0.9179 & 38.4 & 343.3 & 50.8 &   4.0 &  18.97 & 18.32 & 0.25 & $-$0.20 & 0.18 & 0.54 \\
... & ... & ... & ... & ... & ... & ... & ... & ... & ... & ... & ... & ... \\
\hline
\end{tabular}
\end{table*}

\begin{table*}
	\centering
	\caption{Str{\"o}mgren surface magnitudes of the observed sky patches. The layout is analogous to Table~\ref{tab:UBVRIshort}. The full version is available as Table~\ref{tab:uvbylong}.}
	\label{tab:uvbyshort}
\begin{tabular}{lccrrrrrcrrr}
\hline
Sky patch & $D$ & lunar & $z$ & $\gamma$ & $\zeta$ & $\alpha$ & $\theta$ & $\mu(y)$ & $b-y$ & $v-b$ & $u-v$ \\
label &  & phase & ($\degr$) & ($\degr$) & ($\degr$) & ($\degr$) & ($\degr$) & (mag/$\sq\arcsec$) & (mag) & (mag) & (mag) \\
\hline
SP 955    & 232.4330 & 0.9592 & 57.4 &  25.6 & 33.7 & 288.0 &  67.23 & 19.38 & 0.05 & 0.24 & 0.52 \\
SP 4670   & 227.3767 & 0.9182 & 39.3 & 338.9 & 50.7 &   0.4 &  18.90 & 18.27 & 0.20 & 0.31 & 0.72 \\
SP 5524   & 227.3905 & 0.9186 & 41.0 & 331.9 &  7.5 & 341.3 &  33.69 & 19.04 & 0.16 & 0.26 & 0.58 \\
SP 6815   & 232.3318 & 0.9618 & 76.8 &  55.1 & 45.1 &  23.5 &  41.60 & 19.00 & 0.13 & 0.39 & 0.84 \\
SP 7795   & 229.3870 & 0.9915 & 45.6 &   6.7 & 10.1 & 173.1 &  55.49 & 18.94 & 0.06 & 0.22 & 0.46 \\
... & ... & ... & ... & ... & ... & ... & ... & ... & ... & ... & ... \\
\hline
\end{tabular}
\end{table*}

\subsection{Determination of the scattering function}

Computing the scattering function still requires the consideration of the further night sky components. Light pollution is neglected on the basis of the remoteness of the observing site and the deliberate efforts to keep artificial lighting at a minimum in the district. Furthermore, any measurements made during astronomical twilight were not considered in this analysis. It was also decided not to use data points corresponding to lunar zenith angles of $z > 85\degr$ in the determination of the scattering phase function, as these are associated with the greatest uncertainty due to the large airmass and associated extinction of the direct lunar beam as well as Earth curvature effects.

\begin{figure}
	\includegraphics[width=\columnwidth]{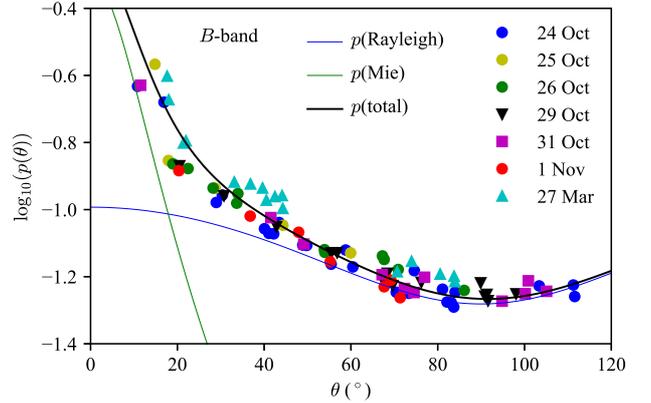}
    \caption{The scattering function in the $B$-band plotted as $\log p(\theta)$ vs. $\theta$. The symbols each represent different observing nights labeled according the date at the start of the night: 24 Oct (lunar phase $\sim 0.78$), 25 Oct (phase $\sim 0.85$), 1 Nov (phase $\sim 0.91$), 26 Oct (phase $\sim 0.92$), 31 Oct (phase $\sim 0.96$), 29 Oct (phase $\sim 0.99$), all in 2012, and 27 Mar 2013 (phase $\sim 1.00$).}
    \label{fig:fphiphase}
\end{figure}

Fig.~\ref{fig:fphiphase} compares the calculated values of $p(\theta)$ obtained for the various nights, and this plot shows no systematic displacement away from this curve for data obtained at specific lunar phases.

\begin{figure}
	\includegraphics[width=\columnwidth]{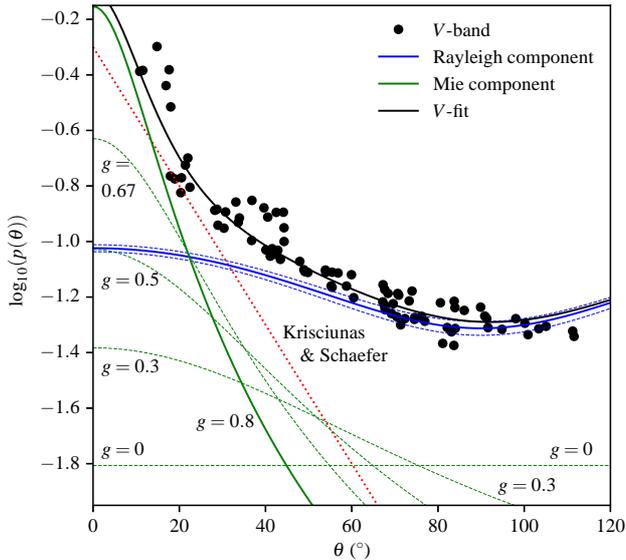}
    \caption{The scattering functions in the $V$-band determined with the single scatter model, plotted as $\log p(\theta)$ vs. $\theta$. The Rayleigh scattering function is shown in blue, the solid line using the adopted $\chi = 0.0148$, and the dashed lines corresponding to $\chi = 0$ and $\chi = 0.0291$. Mie scattering functions with different asymmetries $g$ are plotted in green. The black curve illustrates the fit obtained with $g = 0.8$. The red line, which corresponds to an exponentially declining Mie scattering function as used by \citet{Krisciunas1991}, is unable to match the data as well as a Henyey--Greenstein function.}
    \label{fig:FASS-RM}
\end{figure}

Fig.~\ref{fig:FASS-RM} illustrates attempts at fitting the phase function with different values of $g$. While an asymmetry value of $g > 0.8$ would be able to fit the data even better, it seems unlikely that forward scattering is so pronounced. Atmospheric aerosols typically have values in the range $0.5 < g < 0.7$ \citep{Andrews2006}. It was therefore decided to adopt $g = 0.8$ for the analysis in this paper.

\begin{figure}
	\includegraphics[width=\columnwidth]{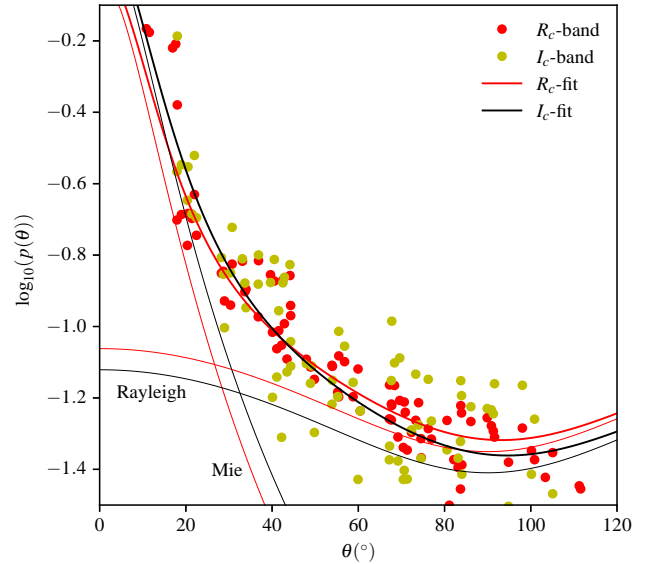}
    \caption{The scattering functions for the $R_c$-band (in red) and $I_c$-band (in yellow), plotted as $\log p(\theta)$ vs. $\theta$, determined with the revised simplified scattering model. The red lines represent the fits to the Rayleigh, Mie and total $R_c$-band scattering functions, while the black curves are the corresponding functions for the $I_c$-band.}
    \label{fig:FASS-RI}
\end{figure}

Fig.~\ref{fig:FASS-RI} highlights how the scatter in the data about the $\log p$ curve progressively grows in the bands most affected by the unpredictable airglow component. This is also evident from the larger standard deviations $\sigma(\delta)$ listed in Table~\ref{tab:filtparms} for the $R_c$ and especially the $I_c$ band.

Table~\ref{tab:examples} explores whether the shift term $\delta$ shows any dependence on the wavelength, scattering angle, lunar and sky patch zenith angles, azimuth or lunar phase. The largest deviation is found for the very small scattering angles ($10\degr$--$20\degr$), the range where the Mie component dominates. A better match for the longer wavelength pass bands could be achieved with an asymmetry parameter $g > 0.8$, or alternatively by the introduction of a scattering angle dependence to $\delta$ in the case of some filters. Another explanation could be an enhanced presence of large particles for which the Henyey-Greenstein function no longer applies, and such an effect would be most noticeable at the longest wavelengths. One also notes a slight discrepancy in Table~\ref{tab:examples} for lunar phases 0.98-1.00, particularly for the $I_c$ band. This may be due the difficulties lunar albedo models have in accurately determining the lunar opposition effect, i.e. the enhanced brightness of the lunar disk at full moon \citep[see, e.g.][]{Krisciunas1991}. No other substantial trends are evident in Table~\ref{tab:examples}.

\begin{table}
	\centering
	\caption{Median of the difference between the model and measured sky brightness (in mag). Positive values indicate a brighter sky than projected by the model.}
	\label{tab:examples}
	\begin{tabular}{lcrrr}
	\hline
		range & no. points & $\Delta U$ & $\Delta V$ & $\Delta I_c$ \\
		\hline
		\multicolumn{5}{l}{{scattering angle $\theta$}} \\
		10$\degr$--20$\degr$ & 8 & $-0.24$ & 0.20 & 0.88 \\
		20$\degr$--60$\degr$ & 39 & $-0.04$ & 0.02 & $-0.01$ \\
		60$\degr$--120$\degr$ & 41 & 0.06 & $-0.05$ & $-0.05$ \\
		\multicolumn{5}{l}{{lunar zenith angle $z$}} \\
		25$\degr$--45$\degr$ & 31 & $-0.03$ & 0.07 & $-0.12$ \\
		45$\degr$--65$\degr$ & 39 & 0.02 & 0.01 & 0.07 \\
		65$\degr$--85$\degr$ & 18 & 0.04 & $-0.07$ & 0.04 \\
		\multicolumn{5}{l}{{sky patch zenith angle $\zeta$}} \\
		0$\degr$--30$\degr$ & 34 & $-0.05$ & $-0.04$ & 0.02 \\
		30$\degr$--45$\degr$ & 26 & 0.04 & 0.02 & $-0.03$ \\
		45$\degr$--65$\degr$ & 28 & 0.03 & 0.01 & 0.05 \\
		\multicolumn{5}{l}{{relative azimuth $|\alpha - \gamma|$}} \\
		0$\degr$--60$\degr$ & 29 & $-0.10$ & 0.02 & 0.19 \\
		60$\degr$--120$\degr$ & 24 & 0.02 & $-0.03$ & $-0.03$ \\
		120$\degr$--180$\degr$ & 35 & 0.04 & 0.01 & $-0.14$ \\
		\multicolumn{5}{l}{{lunar phase}} \\
		0.75-0.90 & 29 & $-0.02$ & $-0.06$ & $-0.20$ \\
		0.90-0.98 & 29 & 0.00 & $-0.04$ & $-0.06$ \\
		0.98-1.00 & 30 & 0.02 & 0.11 & 0.39 \\
		\hline
	\end{tabular}
\end{table}

\subsection{Comparison with the Krisciunas \& Schaefer model}
\label{sec:KScomp}

In order to test the RSS model introduced in Section~\ref{sec:scatter}, the phase function $p(\theta)$ was also determined according to the KS model. As explained in Section~\ref{sec:scatter}, the KS model incorrectly neglects the $(1 - h)$ factor in Equation~\ref{eq:lunarflux}. Without this factor, the relation between the single scattered lunar intensity and the phase function (Equation~\ref{eq:IL1}) would be

\begin{equation}
    I_\text{L1,KS} = p_\text{KS}(\theta) F_{\text{L}}^* \frac{\tau_\text{s}}{\tau} \exp(-\tau \sec{z}) \times (1 - \exp(-\tau \sec{\zeta}) \, .
\end{equation}

\begin{figure}
	\includegraphics[width=\columnwidth]{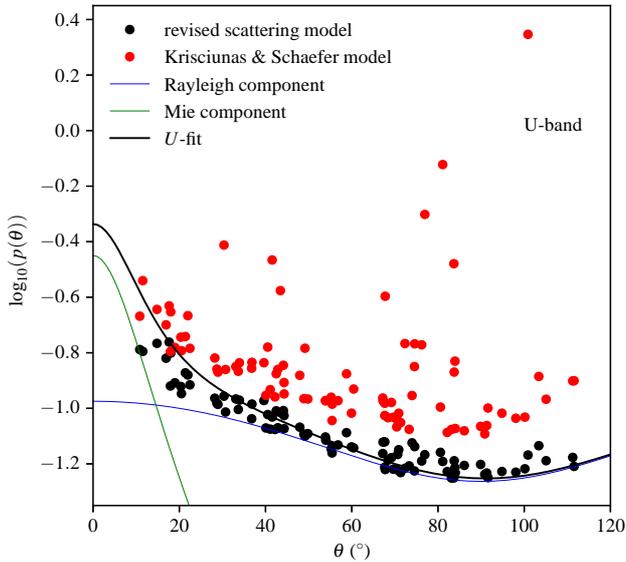}
    \caption{The scattering functions in the $U$-band obtained with the RSS model adopted in this paper (black points) and the \citet{Krisciunas1991} model (KS model, red points), plotted as $\log p(\theta)$ vs. $\theta$.}
    \label{fig:FASSvsKS}
\end{figure}

The RSS and KS phase functions determined using Equation~\ref{eq:IL1} and the above equation respectively are compared for the $U$-band in Fig.~\ref{fig:FASSvsKS}.

It is immediately evident that the scatter is much smaller in this diagram for the points calculated using the RSS model, and that this model results in a much better match with the theoretical phase functions according to equation (\ref{eq:phase}). One notes that the points corresponding to the KS model always lie above the theoretical phase function, sometimes considerably, i.e. the measured sky brightness often far exceeds what is predicted by the KS model. This is to be expected considering that the KS model overestimates the direct lunar beam attenuation, especially for filters like $U$ with large $\tau$ values. Not surprisingly, the points with the largest discrepancy turn out to be those for measurements taken when the moon was less than $10\degr$ above the horizon. While the shift between the two sets of data could in theory be removed by choosing different $\delta$ values, this would not be able to reduce the large scatter for the KS model. Note also that the Mie phase function corresponding to the KS model approach in Fig.~\ref{fig:FASS-RM} was not able to represent $p(\theta)$ well at small scattering angles.

\subsection{The moonlit night sky brightness at SAAO (Sutherland)}

The findings of this investigation allow one to make improved estimates of the sky brightness in the $UBV(RI)_c$ and $uvby$ filters at the Sutherland station of SAAO. The procedure for doing so is summarised below, and can be suitably adapted for use at other observatories.

\noindent - The mean optical depth $\tau$ values for SAAO are listed in Table~\ref{tab:filtparms}, and are usually also well determined at most other observing sites. These constitute fair representations of the atmospheric transparency on most nights, but can also if need be adjusted to suit nights with unusual conditions.

\noindent - The zodiacal light, integrated starlight and airglow contributions to the sky patch intensity are determined using the procedures lined out in Section~\ref{sec:components}. If considered more appropriate, alternative procedures to determine these quantities can be employed here instead. Other sky brightness contributors assumed to be zero in this paper can also be added if required.

\noindent - The top of the atmosphere lunar flux at the time of observation is estimated for the relevant filters. This can be done using the procedure lined out in Section~\ref{sec:lunbright}, or through any other suitable lunar disk brightness calculator.

\noindent - The lunar and sky patch zenith angles determined with astrometric formulae are employed to calculate the gradation function $\Gamma$ as well as the scattering angle $\theta$. The scattering phase function value corresponding to the latter is evaluated with the tabulated Rayleigh and Mie optical depth ratios.

\noindent - The sky intensity due to lunar scattering is now calculated with equation~\ref{eq:logphase}, using the value of the shift $\delta$ from Table~\ref{tab:filtparms} specific to the filter in question.

\noindent - Finally, the sky patch surface brightness $\mu$ is obtained by means of equation~\ref{eq:lunarI}.

\section{Discussion}

\subsection{Performance of the RSS model}

The scatter of the data about the determined offset $\delta$, quantified by the standard deviation $\sigma(\delta)$ (see Table~\ref{tab:filtparms}), is typically only $\sim0.05$ (on a logarithmic scale). The scatter only exceeds this value substantially in the $R_c$ and $I_c$ filters, the bands most affected by the short-term airglow fluctuations.

The comparison of the scattering function determined using the KS formulation and the corresponding data obtained through the RSS model in Fig.~\ref{fig:FASSvsKS} clearly shows the advantage of the approach used in this paper, particularly the fact that the scatter for the KS results is so much larger. An inspection of the assumptions used in the \citeauthor{Krisciunas1991} paper reveals that their direct lunar beam is always considered attenuated by a factor of $\exp(-\tau\sec{z})$ (applying the nomenclature and symbols used in this paper instead of the ones displayed in equation 7 of their paper), i.e. the value appropriate for direct moonlight measured at ground level, even though the scattering event may take place much higher in the atmosphere. The KS model hence underestimates the amount of direct moonlight reaching an arbitrary atmospheric volume element, especially when the volume element is at high altitude. It therefore underestimates the amount of light scattered by this volume element towards the observer as well, and hence also the total singly scattered beam from a particular sky patch, as that is the sum/integral of the light scattered towards the observer by all volume elements in that line of sight. Instead of always equating the direct lunar beam path to the scattering point to the Moon's relative airmass measured from the ground, the RSS model proposed here only considers the lunar beam extinction up to the height of the scattering event. In other words, the direct lunar beam is only attenuated by a factor $\exp(-\tau(1-h)\sec{z})$.

\subsection{Multiple scattering and absorption}

The offset parameter $\delta$ is determined by two ratios related to the atmospheric scattering properties, namely the albedo and the fraction of total to single scattering $\eta$. By definition the albedo is $\le 1$, while $\eta$ must be $\ge 1$, meaning that both positive and negative $\delta$ are possible in principle. It is often assumed that the Rayleigh and Mie optical depths are entirely due to scattering, in which case the albedo would tend towards 1. This implies that a good approximation for the albedo would be $1 - \tau_\text{G} / \tau$. It may however also be that a larger portion of the Mie optical depth than expected is due to absorption. As the Mie component is most noticeable at longer wavelengths, this could lead to greater deviations from the predicted scattering phase function there, particularly in the $I_c$ band.

An excess sky brightness was observed at small scattering angles and longer wavelengths, especially for the $I_c$ band. It suggests the need for an even greater degree of forward scatter than achieved by $g = 0.8$, something not normally observed for atmospheric aerosols. A more likely explanation is that the observed discrepancy is due to particles for which the Henyey-Greenstein scattering function formulation is simply no longer appropriate.

A greater scattering probability translates to a larger fraction of multiply scattered photons. Several previous investigations \citep[e.g.][]{Staude1975} have established that $\eta$ increases with larger $\tau$, and have suggested that $\eta$ can be approximated as $1 + 2.2 \times \tau_\text{R}$ \citep{Noll2012}. This relationship form would however better match the $\delta$ values obtained in the current study and the most likely albedo values if the coefficient of $\tau_\text{R}$ in this expression for $\eta$ is $\approx 4.5$ instead of 2.2.

\subsection{Polarization of moonlight}

The discussion up to this point has ignored one important aspect of moonlight, namely that due to the reflection of the sunlight on the lunar surface, the lunar beam reaching the top of the atmosphere is to some extent polarized. This impacts on the likelihood of a scattering event in specific directions, and hence on the nature of the scattering function. The form of the scattering function presented earlier, in particular its dependence on the scattering angle $\theta$ only, is therefore not completely accurate, and there may be a dependence on the angle of polarization of the incoming beam as well.

A comprehensive study of the lunar beam polarization as a function of lunar phase was carried out by \citet{Dollfus1971}. Their work determined the degree of polarization of the reflected moonlight as a function of lunar phase angle for a variety of lunar surfaces. In particular, they found that the polarization is negligible at all wavelengths for lunar phase angles near $0\degr$ (full moon) and $20\degr$ (lunar phase $\sim0.97$). Between $0\degr$ and $20\degr$ the polarization never significantly exceeds 1 per cent. For lunar phase angles $> 20\degr$ the polarization steadily rises until peaking near $100\degr$ (lunar phase $\sim0.41$). In this regime the polarization also becomes significantly wavelength dependent, with blue light polarization typically approximately double that of red light, and dependent on the nature of the lunar surface.

All observations carried out in this study after 26 October 2012 occurred at lunar phase angles smaller than $35\degr$ (lunar phase $\sim0.91$), when even the blue light polarization is normally small enough to be ignored ($\leq 3$ per cent) in the context of this study. An inspection of figures 2-15 of \citet{Dollfus1971} suggests that for the measurements on 26 October 2012 the lunar phase angle corresponded to a polarization of approximately a third of the maximum value, while on 25 October 2012 about half the maximum polarization would have been achieved. When applying this to the maximum aggregated lunar polarization values calculated through equation 10 of \citeauthor{Dollfus1971} that corresponds to $u$-band polarizations of less than 5 and 7 per cent on 26 and 25 October 2012 respectively, and $V$-band polarizations of $\sim{2.5}$ and $\sim{4}$ per cent on those days. The fact that little or no azimuth dependence is evident in Table~\ref{tab:examples} is consistent with these comparatively low values.

Any bias introduced into this study through the ignoring of the polarization properties of moonlight is thus only expected to have very minor consequences. This view is supported by Fig.~\ref{fig:fphiphase}, which shows no systematic shift between the points representing 25 and 26 October 2012 and those of the remaining days.

\subsection{Sky brightness at SAAO}

South Africa hosts the only large-scale southern hemisphere optical astronomical facilities between the longitudes of South America and Australia, and has as such been a critical site in global monitoring projects. The optimal observing conditions in the country's dry western interior have led to the establishment and continued development of the SAAO facility at Sutherland, which hosts the 11 m South African Large Telescope (SALT) and a wide range of smaller instruments \citep[see, e.g.][]{Catala2013}. As would be expected for a good astronomical site, aerosol presence above SAAO-Sutherland is minimal at most times, although it has been shown to develop to moderate levels following instances of volcanic ash injection into the stratosphere \citep{Kilkenny1995} and the circulation of residue smoke from distant biomass burning \citep{Formenti2002}.

This paper constitutes the first concerted attempt to comprehensively characterise the moonlit sky over SAAO-Sutherland. It provides the framework for determining estimates of the moonlit sky brightness under low aerosol, lunar phase $> 0.75$ conditions that are typically within 0.1 mag of the measured values.

\section{Conclusion}

This work confirms the predicted scattering phase function under moonlit conditions, and the model improvements introduced here produce a far superior match between the observed and fitted scattering function than achieved with the \citeauthor{Krisciunas1991} model. The results confirm that the new approach utilised in the new model leads to a more consistent and also higher projected sky brightness than using the KS formulation, especially when the lunar zenith angle is large. Furthermore, the scattering phase functions obtained here are consistent with the theoretical scattering projections achieved through a straightforward single scattering scenario, with deviations therefrom well explained by multiple scattering.

The study used previously untried procedures to estimate solar beam reflection at the lunar surface as well as the zodiacal light, scattered starlight and airglow contributions, and these achieved satisfactory agreement with the measured sky brightness.

The revised simplified scattering model developed in this study reproduces the lunar sky illumination profile mostly within 0.05 magnitudes at optical wavelengths at the SAAO Sutherland site, and it is expected that this could equally be adapted to other astronomical sites with typically low-to-moderate aerosol loading. While less comprehensive than the sophisticated formulation based on spectral transmission and scattering codes used by \citet{Jones2013}, the revised simplified scattering model presented here offers an effective straightforward way to estimate moonlight sky brightness for optical astronomical observations.

\section*{Acknowledgements}

The author thanks the South African Astronomical Observatory for the allocation of telescope time and use of their facilities.

%%%%%%%%%%%%%%%%%%%%%%%%%%%%%%%%%%%%%%%%%%%%%%%%%%
\section*{Data Availability}

The entire photometric data set for this study is published in Appendix A.



\bibliographystyle{mnras}
\bibliography{SAskybrightness} % if your bibtex file is called example.bib

%\begin{thebibliography}{99}
%\end{thebibliography}

%%%%%%%%%%%%%%%%%%%%%%%%%%%%%%%%%%%%%%%%%%%%%%%%%%

%%%%%%%%%%%%%%%%% APPENDICES %%%%%%%%%%%%%%%%%%%%%
\onecolumn

\appendix

\section{Full tables}

\begin{center}
\begin{longtable}{lrlllrr}
	\caption{List of target stars with associated sky patches.}
	\label{tab:starlist} \\
\hline
RA(2000) & Dec(2000) & Name & MK & Ref & $\Delta$RA & $\Delta$Dec \\
hh\,mm\,ss & $\degr\,\arcmin\,\arcsec$ & & type & & s & \arcmin \\ \hline \endfirsthead

\multicolumn{7}{c}%
{{\tablename\ \thetable{} -- continued from previous page}} \\
\hline 
RA(2000) & Dec(2000) & Name & MK & Ref & $\Delta$RA & $\Delta$Dec \\
hh\,mm\,ss & $\degr\,\arcmin\,\arcsec$ & & type & & s & \arcmin \\ \hline 
\endhead

\hline \multicolumn{7}{r}{{Continued on next page}} \\
\endfoot

\hline
\endlastfoot

00 07 37 & $-$86 02 20 & HD 385    & B9IV     & [W97] &  0 & +1.5 \\
00 13 56 & $-$17 32 43 & HD 955    & B4V      &       & +5 & +1.3 \\
00 15 57 &   +04 15 04 & HD 1160   & A0V      &       & +5 & $-$1.1 \\
00 48 48 &   +18 18 50 & HD 4670   & B9       &       & $-$6 & $-$1.1 \\
00 56 49 & $-$25 21 48 & HD 5524   & A2/3V    &       & +6 & +1.5 \\
01 08 56 &   +09 43 49 & HD 6815   & B9       &       & +4 & +1.3 \\
01 17 02 & $-$42 31 58 & HD 7795   & B9III/IV & E146$^{+}$ &  0 & +2.0 \\
01 31 02 & $-$66 29 48 & HD 9478   & B9V      &       &  0 & +1.6 \\
01 31 30 & $-$44 39 21 & HD 9404   & F3V      & E106$^{+}$ & $-$7 &  0.0 \\
01 31 33 & $-$43 50 47 & HD 9403   & F8V      & E109$^{+}$ & $-$9 &  0.0 \\
01 37 45 & $-$47 10 41 & HD 10101  & G8III    & E170  & $-$13 & 0.0 \\
01 47 09 &   +10 50 39 & HD 10894  & B9       &       & +7 & +2.0 \\
01 49 27 &   +26 28 22 & HD 11079  & B8       &       &  0 & $-$1.3 \\
02 22 46 & $-$14 54 04 & HD 14788  & A1/2IV   &       & +5 & +1.1 \\
02 39 35 &   +01 22 07 & HD 16581  & B9       & [M91] & $-$5 & $-$0.2 \\
02 53 41 & $-$26 09 20 & HD 18100  & B5II/III & [K98] & +5 & $-$1.2 \\
03 14 08 &   +15 35 25 & HD 20086  & A0       &       & +4 & +1.5 \\
03 18 58 & $-$73 59 10 & HD 21166  & F3V      & F102  &  0 & $-$1.6 \\
03 38 40 & $-$75 45 46 & HD 23521  & K2III/IV & F110  & +28 & 0.0 \\
03 46 34 & $-$76 42 57 & HD 24579  & B7III    & [W97] & +17 & +1.5 \\
03 49 49 & $-$42 43 39 & HD 24249  & A5/6V    & E252$^{+}$ & $-$5 & $-$1.0 \\
03 57 52 & $-$46 22 58 & HD 25169  & F6/7V    & E210$^{+}$ & $-$7 & $-$1.3 \\
03 58 34 & $-$23 47 43 & HD 25117  & A0/1V    &       & +4 & +1.3 \\
04 03 29 & $-$44 08 17 & HD 25795  & K3III    & E233  & $-$4 & +1.3 \\
04 13 47 & $-$84 29 09 & HD 29138  & B1Iab    & [W97] &  0 & $-$1.3 \\
04 18 00 & $-$45 39 04 & HD 27471  & G2/3V    & E253$^{+}$ & $-$11 & +1.0 \\
04 30 22 &   +23 35 19 & HD 28482  & B8III    &       &  0 & $-$1.6 \\
05 15 22 &   +00 03 37 & HD 34179  & B8V      & [W90] & +7 & $-$1.1 \\
05 17 45 & $-$33 25 58 & HD 34677  & B9Vp     & [W89] &  0 & +2.0 \\
05 21 22 & $-$14 33 19 & HD 35042  & B5III    & [W97] & +5 & $-$1.0 \\
05 31 05 &   +03 21 11 & HD 36340  & B2V      & [M90] & $-$5 & $-$1.3 \\
06 37 10 & $-$70 55 04 & HD 48467  & B8/9V    &       & +19 & $-$0.4 \\
06 39 46 & $-$43 24 10 & HD 48150  & B3V      & E388$^{+}$ & $-$8 & +0.7 \\
06 45 10 & $-$47 13 22 & HD 49260  & B2III    & E389$^{+}$ & $-$9 & +0.4 \\
06 46 47 & $-$44 58 26 & HD 49559  & K3III    & E338  & $-$2 & $-$1.6 \\
06 48 18 & $-$43 48 04 & HD 49850  & A4V      & E315$^{+}$ &  0 & $-$1.9 \\
06 56 49 &   +20 58 01 & HD 51102  & B8       &       & +6 & 0.0 \\
07 07 24 & $-$14 03 07 & HD 54197  & B2II     & [W97] &  0 & $-$1.4 \\
07 57 04 &   +02 57 03 & HD 65079  & B2Vne    & [L]   & $-$3 & $-$1.5 \\
08 03 52 &   +06 19 25 & HD 66446  & B8       &       & +4 & +1.1 \\
08 04 48 &   +06 11 10 & HD 66665  & B1V      & [M90] & +4 & +1.1 \\
09 13 45 & $-$42 18 37 & HD 79601  & G2V      & E477  & 0 & $-$1.7 \\
09 21 41 & $-$47 19 12 & HD 81035  & A2IV/V   & E410$^{+}$ & $-$6 & +1.8 \\
09 22 01 & $-$46 45 33 & HD 81077  & G9III    & E427  & +9 & +1.5 \\
09 28 40 & $-$45 30 01 & HD 82121  & B5IV/V   & E4100$^{+}$ & $-$2 & $-$1.8 \\
10 33 47 &   +23 21 01 & HD 91427  & A2.5V    &       & +10 & 0.0 \\
10 37 42 & $-$35 43 12 & HD 92136  & B9IV     &       & $-$4 & $-$1.4 \\
10 47 19 &   +06 20 46 & HD 93431  & A4V      &       & +8 & +1.2 \\
12 07 23 & $-$43 14 47 & HD 105283 & F3IV     & E554$^{+}$ & $-$8 & $-$0.3 \\
12 13 50 &   +08 58 20 & HD 106295 & A4Vp     &       &  0 & $-$1.9 \\
12 32 20 & $-$78 11 39 & HD 108927 & B5V      &       & $-$20 & $-$1.0 \\
13 44 31 & $-$17 56 13 & HD 119608 & B1Ib     &       & +3 & $-$1.6 \\
14 18 41 & $-$07 00 49 & HD 125310 & A2IV     &       & +5 & +1.6 \\
14 31 41 & $-$43 05 16 & HD 127294 & F7/8V    & E621$^{+}$ & $-$7 & +0.7 \\
14 40 01 & $-$45 44 35 & HD 128726 & A0V      & E604$^{+}$ & +9 & 0.0 \\
15 08 45 &   +12 29 20 & HD 134305 & ApSrEuCr &       &  0 & +2.2 \\
16 25 02 & $-$50 44 55 & HD 147559 & B9IV     &       & +8 & $-$1.8 \\
17 03 18 & $-$31 36 53 & HD 153855 & B1III    & [K98] &  0 & +1.7 \\
17 20 50 & $-$45 25 11 & HD 156623 & A0V      & E746$^{+}$ & +6 & $-$1.5 \\
17 36 41 & $-$44 52 43 & HD 159384 & K4III    & E739  & $-$9 & +0.6 \\
17 45 57 &   +05 41 40 & HD 161572 & B8       &       & +7 & $-$0.8 \\
18 04 58 & $-$24 40 51 & HD 165016 & B2Ib     & [W97] & +6 & +0.7 \\
18 35 16 & $-$45 56 27 & HD 171141 & B2II/III & [K98] & $-$5 & $-$1.1 \\
19 30 46 & $-$16 10 00 & HD 183570 & B5III    &       &  0 & $-$1.4 \\
20 00 51 & $-$43 58 54 & HD 189247 & F5V      & E813$^{+}$ &  0 & $-$1.3 \\
20 02 10 & $-$44 27 58 & HD 189502 & Ap       & E802$^{+}$ & +4 & +1.3 \\
20 08 30 & $-$74 45 27 & HD 189723 & K0III    & F315  &  0 & +1.3 \\
20 13 53 & $-$45 09 50 & HD 191849 & M0V      & E861  & +1 & +1.8 \\
20 18 16 & $-$42 51 36 & HD 192758 & F0V      & E810$^{+}$ & $-$6 & $-$1.1 \\
20 43 13 & $-$76 32 44 & HD 196227 & G2V      & F323  & +19 & +1.0 \\
21 27 23 & $-$13 35 15 & HD 204220 & B9III/IV &       & +6 & +0.5 \\
21 35 55 &   +05 28 35 & HD 205556 & B9       & [M91] &  0 & +1.9 \\
21 59 18 & $-$23 49 56 & HD 208793 & B9V      &       & $-$5 & $-$1.0 \\
22 08 12 & $-$03 31 53 & HD 210121 & B3V      &       & $-$6 & +1.0 \\
22 37 19 & $-$44 39 54 & HD 214174 & K1III    & E932  &  0 & +1.6 \\
22 38 04 & $-$46 42 45 & HD 214308 & F5V      & E977$^{+}$ & +9 & +1.0 \\
22 41 26 &   +23 50 48 & HD 214930 & B2IV     &       & +9 & +0.7 \\
22 45 38 &   +03 37 52 & HD 215512 & B9       &       & $-$6 & +0.7 \\
22 47 26 & $-$44 57 54 & HD 215657 & G3V      & E917$^{+}$ & +7 & +1.4 \\
22 49 51 & $-$44 25 25 & HD 216009 & A0V      & E901$^{+}$ & +6 & +1.7 \\
23 02 02 & $-$59 27 46 & HD 217505 & B2III/IV & [K98] & +12 & +0.9 \\
23 17 28 & $-$16 10 20 & HD 219639 & B5II/III & [W97] & $-$6 & +1.5 \\
23 21 51 & $-$09 45 41 & HD 220172 & B3Vn     & [K98] & +7 & 0.0 \\
\hline

\multicolumn{7}{l}{Notes:} \\
\multicolumn{7}{l}{$^{+}$These stars are also Str{\"o}mgren photometry standards from } \\
\multicolumn{7}{l}{\citet{Kilkenny1992}} \\
\multicolumn{7}{l}{[K98] - \citet{Kilkenny1998}, [M90] - \citet{Menzies1990},} \\
\multicolumn{7}{l}{[M91] - \citet{Menzies1991}, [W89] - \citet{Winkler1989},} \\
\multicolumn{7}{l}{[W90] - \citet{Winkler1990}, [W97] - \citet{Winkler1997}}

\end{longtable}

\newpage
\begin{table}
	\centering
	\caption{$UBV(RI)_c$ magnitudes of observed stars. The second column lists the mid-point in time of each observation $D$, in fractional days, relative to the heliocentric Julian day (HJD) 2456000.0, i.e. $D = \mbox{HJD} - 2456000$.}
	\label{tab:JClong}
\begin{tabular}{lccrrrr}
\hline
Name & $D$ & $V$ & $B-V$ & $U-B$ & $V-R_c$ & $V-I_c$ \\
\hline
HD 385    & 225.384 & 7.40 & $-$0.01 & $-$0.16 &    0.00 &    0.01 \\
HD 955    & 225.498 & 7.36 & $-$0.16 & $-$0.64 & $-$0.08 & $-$0.16 \\
          & 232.428 & 7.39 & $-$0.14 & $-$0.64 & $-$0.08 & $-$0.16 \\
HD 1160   & 226.342 & 7.10 &    0.04 &    0.04 &    0.00 &    0.01 \\
HD 4670   & 227.369 & 7.91 &    0.00 & $-$0.16 & $-$0.01 &    0.00 \\
HD 5524   & 227.385 & 7.21 &    0.11 &    0.10 &    0.05 &    0.10 \\
HD 6815   & 225.490 & 7.29 & $-$0.05 & $-$0.24 & $-$0.04 & $-$0.07 \\
          & 232.326 & 7.29 & $-$0.06 & $-$0.22 & $-$0.04 & $-$0.07 \\
HD 9478   & 225.394 & 8.31 & $-$0.06 & $-$0.25 & $-$0.04 & $-$0.07 \\
HD 10894  & 227.415 & 7.04 & $-$0.02 & $-$0.10 & $-$0.01 & $-$0.02 \\
HD 11079  & 225.444 & 6.88 & $-$0.06 & $-$0.48 & $-$0.04 & $-$0.05 \\
HD 14788  & 225.436 & 7.68 &    0.03 &    0.06 &    0.02 &    0.04 \\
          & 229.430 & 7.68 &    0.03 &    0.07 &    0.02 &    0.03 \\
HD 16581  & 229.444 & 8.19 & $-$0.06 & $-$0.29 & $-$0.03 & $-$0.06 \\
HD 20086  & 225.481 & 7.15 &    0.06 &    0.04 &    0.02 &    0.04 \\
          & 232.380 & 7.16 &    0.07 &    0.05 &    0.02 &    0.05 \\
HD 24579  & 225.403 & 8.06 & $-$0.02 & $-$0.39 & $-$0.01 & $-$0.01 \\
HD 25117  & 225.551 & 7.91 &    0.03 &    0.02 &    0.00 &    0.01 \\
HD 28482  & 225.561 & 7.15 &    0.41 &    0.02 &    0.24 &    0.56 \\
          & 229.521 & 7.15 &    0.42 &    0.01 &    0.24 &    0.56 \\
HD 29138  & 232.400 & 7.22 & $-$0.06 & $-$0.84 & $-$0.01 & $-$0.03 \\
HD 34179  & 229.462 & 8.04 & $-$0.05 & $-$0.43 &    0.00 & $-$0.02 \\
HD 34677  & 233.563 & 7.86 & $-$0.07 & $-$0.23 & $-$0.03 & $-$0.06 \\
HD 35042  & 225.454 & 7.29 & $-$0.09 & $-$0.56 & $-$0.04 & $-$0.08 \\
          & 233.511 & 7.30 & $-$0.09 & $-$0.56 & $-$0.04 & $-$0.08 \\
HD 36340  & 233.529 & 7.97 & $-$0.14 & $-$0.77 & $-$0.07 & $-$0.14 \\
HD 48467  & 229.505 & 8.28 & $-$0.06 & $-$0.30 & $-$0.03 & $-$0.06 \\
          & 379.260 & 8.27 & $-$0.06 & $-$0.30 & $-$0.04 & $-$0.06 \\
HD 51102  & 229.576 & 7.41 & $-$0.10 & $-$0.43 & $-$0.06 & $-$0.09 \\
HD 54197  & 229.562 & 7.99 & $-$0.06 & $-$0.78 &    0.00 & $-$0.02 \\
HD 65079  & 225.573 & 7.82 & $-$0.15 & $-$0.77 & $-$0.06 & $-$0.11 \\
HD 66446  & 233.581 & 7.78 &    0.29 &    0.12 &    0.16 &    0.33 \\
HD 66665  & 225.584 & 7.82 & $-$0.24 & $-$0.99 & $-$0.13 & $-$0.26 \\
HD 91427  & 379.319 & 7.31 &    0.22 &    0.10 &    0.14 &    0.26 \\
HD 92136  & 379.282 & 6.98 &    0.03 & $-$0.01 &    0.02 &    0.05 \\
HD 93431  & 379.412 & 7.11 &    0.17 &    0.10 &    0.05 &    0.13 \\
HD 106295 & 379.336 & 7.58 &    0.25 &    0.08 &    0.11 &    0.24 \\
HD 108927 & 379.387 & 7.77 &    0.09 & $-$0.33 &    0.06 &    0.13 \\
HD 119608 & 379.369 & 7.52 & $-$0.06 & $-$0.82 & $-$0.02 & $-$0.04 \\
HD 125310 & 379.550 & 7.42 &    0.15 &    0.10 &    0.09 &    0.18 \\
HD 134305 & 379.567 & 7.25 &    0.21 &    0.12 &    0.10 &    0.20 \\
HD 147559 & 379.529 & 7.88 &    0.07 & $-$0.04 &    0.03 &    0.08 \\
HD 161572 & 379.606 & 7.55 &    0.00 & $-$0.46 &    0.00 &    0.02 \\
HD 165016 & 379.633 & 7.30 & $-$0.03 & $-$0.85 &    0.01 &    0.02 \\
HD 183570 & 227.275 & 7.41 & $-$0.01 & $-$0.44 &    0.01 &    0.02 \\
HD 204220 & 232.300 & 7.08 & $-$0.09 & $-$0.44 & $-$0.04 & $-$0.08 \\
HD 205556 & 227.288 & 8.29 & $-$0.06 & $-$0.37 & $-$0.03 & $-$0.05 \\
HD 208793 & 226.315 & 7.01 &    0.01 &    0.02 &    0.01 &    0.01 \\
HD 210121 & 226.328 & 7.63 &    0.17 & $-$0.29 &    0.10 &    0.21 \\
HD 214930 & 225.333 & 7.36 & $-$0.12 & $-$0.65 & $-$0.07 & $-$0.12 \\
HD 215512 & 225.324 & 7.90 & $-$0.05 & $-$0.42 & $-$0.01 & $-$0.03 \\
          & 232.313 & 7.87 & $-$0.05 & $-$0.42 & $-$0.02 & $-$0.04 \\
HD 219639 & 227.319 & 6.69 & $-$0.14 & $-$0.58 & $-$0.07 & $-$0.14 \\
\hline
\end{tabular}
\end{table}

\begin{table}
    \centering
	\caption{Results of the Str{\"o}mgren photometry of the observed stars. As in Table~\ref{tab:JClong}, the second column lists $D = \mbox{HJD} - 2456000$. In line with common practice, the $u$ and $v$ magnitudes are not given explicitly, but rather in terms of the indexes $m_1 = (v-b) - (b-y)$ measuring metallicity and $c_1 = (u-v) - (v-b)$ quantifying the Balmer jump.}
	\label{tab:stromlong}
\begin{tabular}{lccrrr}
\hline
Name & $D$ & $y$ & $b-y$ & $m_1$ & $c_1$ \\
\hline
HD 955    & 232.433 & 7.43 & $-$0.07 & 0.10 & 0.34 \\
HD 1160   & 226.348 & 7.11 &    0.00 & 0.19 & 0.99 \\
HD 4670   & 227.376 & 7.94 &    0.02 & 0.11 & 0.92 \\
HD 5524   & 227.390 & 7.22 &    0.05 & 0.19 & 1.05 \\
HD 6815   & 232.332 & 7.30 & $-$0.03 & 0.12 & 0.84 \\
HD 10101  & 232.356 & 7.59 &    0.62 & 0.38 & 0.34 \\
HD 10894  & 227.421 & 7.07 &    0.00 & 0.14 & 1.01 \\
HD 14788  & 229.436 & 7.69 &    0.02 & 0.15 & 1.22 \\
HD 16581  & 229.450 & 8.19 & $-$0.02 & 0.11 & 0.74 \\
HD 18100  & 232.369 & 8.51 & $-$0.10 & 0.08 & 0.00 \\
HD 20086  & 232.386 & 7.17 &    0.04 & 0.18 & 1.04 \\
HD 21166  & 232.419 & 7.28 &    0.28 & 0.12 & 0.52 \\
HD 23521  & 229.422 & 7.67 &    0.71 & 0.66 & 0.29 \\
HD 28482  & 229.526 & 7.14 &    0.33 & 0.00 & 0.87 \\
HD 29138  & 232.405 & 7.22 &    0.04 & 0.01 & $-$0.01 \\
HD 34179  & 229.468 & 8.05 & $-$0.01 & 0.10 & 0.55 \\
HD 34677  & 233.568 & 7.86 & $-$0.03 & 0.12 & 0.88 \\
HD 35042  & 233.516 & 7.30 & $-$0.02 & 0.10 & 0.35 \\
HD 36340  & 233.524 & 7.97 & $-$0.05 & 0.07 & 0.14 \\
HD 48467  & 229.510 & 8.28 & $-$0.02 & 0.12 & 0.73 \\
          & 379.266 & 8.26 & $-$0.02 & 0.12 & 0.74 \\
HD 49559  & 233.542 & 7.97 &    0.92 & 0.71 & 0.35 \\
HD 51102  & 229.582 & 7.41 & $-$0.04 & 0.10 & 0.64 \\
HD 54197  & 229.568 & 8.00 &    0.02 & 0.05 & 0.04 \\
HD 66446  & 233.576 & 7.77 &    0.16 & 0.26 & 0.72 \\
HD 79601  & 233.550 & 8.02 &    0.38 & 0.14 & 0.30 \\
HD 91427  & 379.324 & 7.28 &    0.14 & 0.16 & 0.94 \\
HD 92136  & 379.276 & 6.98 &    0.02 & 0.12 & 1.16 \\
HD 93431  & 379.406 & 7.12 &    0.07 & 0.21 & 0.95 \\
HD 97991  & 379.297 & 7.38 & $-$0.09 & 0.09 & 0.01 \\
HD 106295 & 379.331 & 7.58 &    0.14 & 0.17 & 0.90 \\
HD 108927 & 379.382 & 7.76 &    0.10 & 0.06 & 0.50 \\
HD 125310 & 379.555 & 7.43 &    0.09 & 0.16 & 1.04 \\
HD 134305 & 379.562 & 7.24 &    0.11 & 0.20 & 0.97 \\
HD 147559 & 379.524 & 7.89 &    0.07 & 0.08 & 1.08 \\
HD 153855 & 379.653 & 6.98 &    0.02 & 0.01 & $-$0.07 \\
HD 159384 & 379.585 & 7.36 &    0.87 & 0.78 & 0.34 \\
HD 161572 & 379.612 & 7.56 &    0.04 & 0.06 & 0.43 \\
HD 165016 & 379.628 & 7.31 &    0.06 & 0.00 & $-$0.06 \\
HD 171141 & 379.592 & 8.39 & $-$0.10 & 0.08 & $-$0.02 \\
HD 183570 & 227.280 & 7.40 &    0.04 & 0.06 & 0.51 \\
HD 189723 & 227.354 & 7.29 &    0.65 & 0.48 & 0.33 \\
HD 196227 & 232.280 & 7.65 &    0.38 & 0.20 & 0.42 \\
HD 204220 & 232.305 & 7.09 & $-$0.04 & 0.11 & 0.54 \\
HD 205556 & 227.294 & 8.29 & $-$0.01 & 0.10 & 0.63 \\
HD 208793 & 226.320 & 7.02 &    0.01 & 0.17 & 1.07 \\
HD 210121 & 226.333 & 7.64 &    0.17 & 0.06 & 0.43 \\
HD 214174 & 225.272 & 7.94 &    0.77 & 0.57 & 0.28 \\
HD 215512 & 232.318 & 7.89 &    0.00 & 0.09 & 0.54 \\
HD 219639 & 227.325 & 6.71 & $-$0.06 & 0.09 & 0.44 \\
\hline
\end{tabular}
\end{table}
\clearpage

\begin{longtable}{lccrrrrrrrrrr}
	\caption{$UBV(RI)_c$ surface magnitudes of the observed sky patches. Columns 3-8 respectively list the lunar phase, zenith angle and azimuth, sky patch zenith angle and azimuth, and scattering angle at the time of observation.}
	\label{tab:UBVRIlong} \\
\hline
Sky patch & $D$ & lunar & $z$ & $\gamma$ & $\zeta$ & $\alpha$ & $\theta$ & $\mu(V)$ & $B-V$ & $U-B$ & $V-R_c$ & $V-I_c$ \\
label &  & phase & ($\degr$) & ($\degr$) & ($\degr$) & ($\degr$) & ($\degr$) & (mag/$\sq\arcsec$) & (mag) & (mag) & (mag) & (mag) \\
\hline \endfirsthead

\multicolumn{13}{c}%
{{\tablename\ \thetable{} -- continued from previous page}} \\
\hline
Sky patch & $D$ & lunar & $z$ & $\gamma$ & $\zeta$ & $\alpha$ & $\theta$ & $\mu(V)$ & $B-V$ & $U-B$ & $V-R_c$ & $V-I_c$ \\
label &  & phase & ($\degr$) & ($\degr$) & ($\degr$) & ($\degr$) & ($\degr$) & (mag/$\sq\arcsec$) & (mag) & (mag) & (mag) & (mag) \\
\hline \endhead

\hline \multicolumn{13}{r}{{Continued on next page}} \\
\endfoot

\hline
\endlastfoot

SP 385    & 225.3840 & 0.7745 & 45.2 & 300.5 & 53.6 & 180.9 &  82.15 & 19.87 & 0.16 & $-$0.38 & 0.18 & 0.99 \\
SP 955    & 225.4980 & 0.7823 & 77.7 & 273.4 & 47.4 & 275.7 &  30.36 & 19.48 & 0.31 &  0.01 & 0.20 & 0.85 \\
          & 232.4279 & 0.9594 & 58.0 &  27.4 & 32.3 & 289.8 &  67.19 & 19.47 & 0.05 & $-$0.38 & 0.02 & 0.47 \\
SP 1160   & 226.3422 & 0.8527 & 34.4 & 338.6 & 37.0 &   9.2 &  17.94 & 18.77 & 0.25 & $-$0.20 & 0.14 & 0.55 \\
SP 4670   & 227.3685 & 0.9179 & 38.4 & 343.3 & 50.8 &   4.0 &  18.97 & 18.32 & 0.25 & $-$0.20 & 0.18 & 0.54 \\
SP 5524   & 227.3853 & 0.9185 & 40.3 & 334.4 &  7.1 & 354.3 &  33.68 & 19.12 & 0.15 & $-$0.33 & 0.07 & 0.37 \\
SP 6815   & 225.4901 & 0.7817 & 75.4 & 274.9 & 53.0 & 316.9 &  43.46 & 19.53 & 0.25 & $-$0.06 & 0.18 & 0.83 \\
          & 232.3266 & 0.9620 & 78.1 &  56.3 & 45.7 &  25.9 &  41.55 & 19.04 & 0.24 &  0.02 & 0.14 & 0.59 \\
SP 7795   & 229.3814 & 0.9915 & 45.8 &   9.4 & 10.5 & 164.9 &  55.47 & 19.03 & 0.03 & $-$0.40 & 0.04 & 0.38 \\
SP 9403   & 232.3393 & 0.9616 & 75.0 &  53.3 & 19.0 & 132.4 &  72.33 & 19.82 & 0.11 & $-$0.20 & 0.12 & 0.73 \\
SP 9404   & 227.4005 & 0.9190 & 42.6 & 327.1 & 12.4 & 172.2 &  53.98 & 19.50 & 0.10 & $-$0.41 & 0.06 & 0.47 \\
SP 9478   & 225.3935 & 0.7751 & 47.7 & 297.4 & 34.3 & 175.2 &  70.43 & 20.12 & 0.15 & $-$0.39 & 0.14 & 0.94 \\
SP 10101  & 232.3506 & 0.9613 & 72.5 &  50.6 & 19.6 & 143.7 &  74.54 & 19.80 & 0.10 & $-$0.25 & 0.09 & 0.65 \\
SP 10894  & 227.4149 & 0.9195 & 45.2 & 320.8 & 43.3 &   1.6 &  28.23 & 18.70 & 0.17 & $-$0.25 & 0.09 & 0.40 \\
SP 11079  & 225.4440 & 0.7783 & 61.8 & 284.3 & 59.2 & 352.9 &  58.78 & 19.48 & 0.16 & $-$0.22 & 0.16 & 0.92 \\
SP 14788  & 225.4361 & 0.7778 & 59.5 & 286.1 & 18.0 &  13.7 &  60.44 & 20.11 & 0.15 & $-$0.36 & 0.17 & 0.91 \\
          & 229.4305 & 0.9920 & 46.6 & 345.9 & 17.7 &   7.7 &  30.73 & 18.47 & 0.13 & $-$0.28 & 0.09 & 0.34 \\
SP 16581  & 229.4438 & 0.9921 & 47.8 & 339.9 & 33.8 &   3.2 &  20.50 & 18.04 & 0.21 & $-$0.19 & 0.13 & 0.37 \\
SP 18100  & 232.3637 & 0.9610 & 69.5 &  47.2 & 27.9 &  85.2 &  49.15 & 19.34 & 0.11 & $-$0.22 & 0.07 & 0.50 \\
SP 20086  & 225.4813 & 0.7810 & 72.8 & 276.6 & 48.0 &   1.2 &  74.60 & 19.94 & 0.16 & $-$0.19 & 0.19 & 0.98 \\
          & 232.3800 & 0.9605 & 66.2 &  42.7 & 56.2 &  36.1 &  11.53 & 17.19 & 0.68 &  0.26 & 0.45 & 1.08 \\
SP 21166  & 225.6087 & 0.7922 &   -  &   -   & 47.0 & 195.2 &  82.17 & 21.10 & 0.08 & $-$0.21 & 0.59 & 1.91 \\
          & 232.4141 & 0.9597 & 60.0 &  32.1 & 42.7 & 172.3 &  94.83 & 19.56 & 0.04 & $-$0.38 & 0.03 & 0.60 \\
SP 23521  & 229.4155 & 0.9918 & 45.7 & 353.0 & 45.2 & 171.0 &  90.86 & 19.04 & 0.01 & $-$0.41 & 0.02 & 0.47 \\
SP 24249  & 225.5154 & 0.7837 & 82.9 & 270.2 & 10.5 & 190.6 &  81.10 & 20.68 & 0.30 & $-$0.10 & 0.30 & 1.29 \\
SP 24579  & 225.4030 & 0.7756 & 50.2 & 294.6 & 47.4 & 169.1 &  84.01 & 20.00 & 0.09 & $-$0.32 & 0.17 & 1.03 \\
SP 25117  & 225.5511 & 0.7867 & 93.5 & 263.7 & 14.5 & 303.2 &  82.36 & 21.63 & 0.76 & $-$0.60 & 0.67 & 2.08 \\
SP 25169  & 225.4134 & 0.7763 & 53.1 & 291.7 & 30.9 & 127.6 &  83.13 & 20.24 & 0.13 & $-$0.38 & 0.18 & 1.06 \\
          & 229.5914 & 0.9937 & 78.9 & 294.1 & 28.1 & 230.8 &  67.76 & 19.40 & 0.15 & $-$0.07 & 0.16 & 0.86 \\
SP 25795  & 225.5079 & 0.7826 & 80.6 & 271.6 & 12.0 & 167.8 &  83.69 & 20.61 & 0.32 & $-$0.14 & 0.30 & 1.31 \\
SP 27471  & 229.4925 & 0.9926 & 55.2 & 320.6 & 14.8 & 155.5 &  69.61 & 19.25 & 0.03 & $-$0.40 & 0.06 & 0.54 \\
          & 233.4986 & 0.9117 & 55.0 &  13.0 & 13.4 & 172.2 &  67.65 & 19.89 & 0.12 & $-$0.38 & 0.10 & 0.82 \\
SP 28482  & 225.5607 & 0.7877 &   -  &   -   & 56.6 & 350.4 &  92.28 & 20.96 & 0.85 & $-$0.34 & 0.70 & 2.00 \\
          & 229.5211 & 0.9929 & 61.2 & 311.5 & 56.0 &   1.8 &  42.85 & 18.28 & 0.11 & $-$0.18 & 0.09 & 0.48 \\
SP 29138  & 232.4003 & 0.9600 & 62.3 &  36.6 & 53.3 & 175.8 & 105.07 & 19.32 & 0.03 & $-$0.33 & 0.07 & 0.62 \\
SP 34179  & 229.4623 & 0.9923 & 50.1 & 332.0 & 45.7 &  51.6 &  56.81 & 18.66 & 0.06 & $-$0.33 & 0.04 & 0.36 \\
SP 34677  & 233.5629 & 0.9097 & 55.0 & 347.2 &  4.8 & 256.4 &  55.23 & 19.76 & 0.11 & $-$0.36 & 0.09 & 0.65 \\
SP 35042  & 225.4540 & 0.7790 & 64.7 & 282.1 & 42.5 &  75.5 & 103.38 & 20.08 & 0.11 & $-$0.41 & 0.17 & 0.98 \\
          & 233.5107 & 0.9113 & 54.4 &   8.2 & 22.0 &  39.2 &  36.78 & 19.33 & 0.14 & $-$0.29 & 0.11 & 0.61 \\
SP 36340  & 233.5294 & 0.9108 & 54.0 &   0.6 & 36.9 &  16.5 &  20.35 & 18.80 & 0.21 & $-$0.21 & 0.14 & 0.56 \\
SP 48150  & 225.4625 & 0.7797 & 67.2 & 280.3 & 46.8 & 121.4 & 111.25 & 20.02 & 0.12 & $-$0.28 & 0.17 & 1.04 \\
          & 229.3972 & 0.9916 & 45.4 &   1.8 & 60.7 & 125.3 &  89.89 & 18.63 & 0.02 & $-$0.32 & 0.00 & 0.42 \\
          & 233.5888 & 0.9089 & 57.4 & 337.4 & 11.8 & 160.3 &  69.19 & 19.91 & 0.07 & $-$0.39 & 0.12 & 0.76 \\
SP 48467  & 229.5047 & 0.9927 & 57.6 & 316.5 & 43.9 & 162.7 &  98.05 & 19.08 & 0.00 & $-$0.37 & 0.04 & 0.58 \\
          & 379.2604 & 0.9975 & 70.0 &  85.6 & 40.1 & 190.2 &  83.77 & 18.79 & 0.03 & $-$0.25 & 0.00 & 0.38 \\
SP 49260  & 225.6009 & 0.7914 &   -  &   -   & 16.8 & 155.1 & 109.81 & 21.60 & 0.50 & $-$0.71 & 0.65 & 1.98 \\
          & 229.5349 & 0.9930 & 64.4 & 307.5 & 27.3 & 131.8 &  91.60 & 19.35 & 0.00 & $-$0.35 & 0.08 & 0.68 \\
SP 49559  & 225.4703 & 0.7802 & 69.5 & 278.8 & 45.9 & 123.5 & 111.56 & 20.08 & 0.18 & $-$0.26 & 0.20 & 1.10 \\
          & 233.5370 & 0.9105 & 54.0 & 357.6 & 23.7 & 129.7 &  71.34 & 19.90 & 0.10 & $-$0.39 & 0.14 & 0.80 \\
SP 49850  & 225.5928 & 0.7906 &   -  &   -   & 15.9 & 140.2 & 111.96 & 21.67 & 0.60 & $-$0.59 & 0.64 & 1.93 \\
          & 229.4791 & 0.9924 & 52.8 & 325.5 & 41.2 & 121.4 &  91.31 & 19.10 & 0.00 & $-$0.39 & 0.03 & 0.50 \\
          & 379.2509 & 0.9976 & 72.8 &  87.4 & 16.0 & 220.2 &  83.94 & 19.06 & 0.04 & $-$0.22 & 0.04 & 0.41 \\
SP 51102  & 229.5766 & 0.9935 & 74.9 & 297.3 & 56.1 &  20.6 &  76.22 & 18.88 & 0.10 & $-$0.10 & 0.10 & 0.71 \\
SP 54197  & 229.5626 & 0.9933 & 71.2 & 300.5 & 29.8 &  58.3 &  86.60 & 19.36 & 0.05 & $-$0.26 & 0.12 & 0.79 \\
SP 65079  & 225.5729 & 0.7888 &   -  &   -   & 50.7 &  53.4 & 142.54 & 21.82 & 0.76 & $-$0.65 & 0.69 & 2.10 \\
SP 66446  & 233.5813 & 0.9092 & 56.6 & 340.1 & 47.7 &  41.2 &  47.95 & 19.18 & 0.13 & $-$0.25 & 0.09 & 0.58 \\
SP 66665  & 225.5838 & 0.7898 &   -  &   -   & 51.9 &  48.8 & 144.31 & 21.28 & 0.80 & $-$0.57 & 0.69 & 2.08 \\
SP 79601  & 225.5413 & 0.7859 & 90.2 & 265.5 & 54.3 & 121.8 & 130.97 & 20.96 & 0.70 & $-$0.14 & 0.61 & 1.77 \\
          & 233.5555 & 0.9100 & 54.6 & 350.1 & 44.8 & 119.6 &  87.38 & 19.68 & 0.10 & $-$0.36 & 0.15 & 0.88 \\
SP 81035  & 225.5247 & 0.7845 & 85.6 & 268.5 & 58.5 & 129.3 & 127.13 & 20.16 & 0.44 &  0.13 & 0.36 & 1.31 \\
          & 379.3953 & 0.9962 & 32.9 &  46.4 & 26.0 & 226.6 &  58.88 & 18.62 & 0.00 & $-$0.45 & $-$0.04 & 0.01 \\
SP 81077  & 379.3103 & 0.9970 & 55.3 &  75.2 & 14.6 & 172.5 &  58.38 & 18.62 & 0.06 & $-$0.34 & 0.01 & 0.19 \\
SP 82121  & 229.5487 & 0.9932 & 67.7 & 303.9 & 51.6 & 125.2 & 119.29 & 18.84 & 0.03 & $-$0.25 & 0.05 & 0.63 \\ 
SP 91427  & 379.3189 & 0.9970 & 52.8 &  73.1 & 58.1 &  19.1 &  44.15 & 17.63 & 0.16 & $-$0.16 & 0.02 & 0.10 \\
SP 92136  & 379.2820 & 0.9973 & 63.6 &  81.3 & 26.5 & 106.3 &  40.52 & 18.19 & 0.14 & $-$0.20 & 0.04 & 0.19 \\
SP 93431  & 379.4115 & 0.9961 & 29.6 &  37.5 & 40.4 & 340.8 &  33.12 & 17.89 & 0.09 & $-$0.31 & 0.01 & 0.02 \\
SP 97991  & 379.2916 & 0.9972 & 60.8 &  79.3 & 45.8 &  59.0 &  21.98 & 17.43 & 0.22 & $-$0.09 & 0.08 & 0.23 \\
SP 105283 & 379.3433 & 0.9967 & 46.0 &  66.4 & 27.4 & 122.8 &  36.85 & 18.01 & 0.12 & $-$0.26 & 0.00 & 0.04 \\
SP 106295 & 379.3360 & 0.9968 & 48.0 &  68.6 & 53.9 &  46.6 &  18.01 & 16.81 & 0.35 & $-$0.01 & 0.22 & 0.50 \\
SP 108927 & 379.3870 & 0.9963 & 34.7 &  50.4 & 46.9 & 173.9 &  70.65 & 18.54 & $-$0.02 & $-$0.44 & $-$0.06 & 0.00 \\
SP 119608 & 379.3687 & 0.9965 & 39.2 &  58.0 & 44.8 &  83.2 &  17.64 & 16.66 & 0.48 &  0.00 & 0.30 & 0.67 \\
SP 125310 & 379.5501 & 0.9947 & 37.5 & 302.8 & 26.8 & 338.5 &  21.41 & 17.80 & 0.13 & $-$0.26 & $-$0.03 & $-$0.04 \\
SP 127294 & 379.3613 & 0.9966 & 41.2 &  60.7 & 48.9 & 121.4 &  42.49 & 17.82 & 0.13 & $-$0.26 & $-$0.01 & 0.02 \\
SP 128726 & 379.5124 & 0.9951 & 29.3 & 321.6 & 15.3 & 154.4 &  44.30 & 18.37 & 0.06 & $-$0.38 & $-$0.03 & $-$0.06 \\
SP 134305 & 379.5669 & 0.9946 & 41.8 & 296.6 & 44.9 & 355.6 &  39.64 & 17.97 & 0.10 & $-$0.29 & $-$0.02 & $-$0.02 \\
SP 147559 & 379.5290 & 0.9950 & 32.6 & 312.3 & 28.5 & 139.1 &  61.03 & 18.63 & 0.02 & $-$0.44 & $-$0.03 & 0.09 \\
SP 156623 & 379.6403 & 0.9936 & 62.4 & 277.3 & 13.4 & 169.5 &  67.20 & 18.93 & 0.05 & $-$0.33 & 0.04 & 0.36 \\
SP 159384 & 379.5802 & 0.9944 & 45.4 & 292.3 & 25.7 & 127.5 &  70.42 & 18.71 & 0.07 & $-$0.40 & 0.04 & 0.27 \\
SP 161572 & 379.6065 & 0.9941 & 52.7 & 285.1 & 43.3 &  32.7 &  73.98 & 18.71 & $-$0.03 & $-$0.41 & $-$0.05 & 0.19 \\
SP 165016 & 379.6331 & 0.9937 & 60.3 & 278.9 & 16.8 &  67.1 &  74.91 & 18.99 & 0.03 & $-$0.41 & 0.06 & 0.40 \\
SP 171141 & 379.5973 & 0.9942 & 50.1 & 287.5 & 31.7 & 126.4 &  80.57 & 18.92 & $-$0.04 & $-$0.46 & $-$0.05 & 0.22 \\
SP 183570 & 227.2747 & 0.9148 & 42.7 &  34.9 & 41.5 & 282.6 &  67.65 & 19.28 & 0.10 & $-$0.37 & 0.08 & 0.43 \\
SP 189247 & 226.2986 & 0.8508 & 32.5 &   6.5 & 34.9 & 238.1 &  59.93 & 19.54 & 0.15 & $-$0.33 & 0.13 & 0.50 \\
SP 189502 & 225.3467 & 0.7723 & 36.2 & 315.3 & 46.3 & 237.3 &  49.81 & 19.60 & 0.16 & $-$0.35 & 0.12 & 0.72 \\
          & 227.3052 & 0.9158 & 38.6 &  19.5 & 37.1 & 237.6 &  70.89 & 19.45 & 0.08 & $-$0.41 & 0.04 & 0.38 \\
SP 189723 & 227.3480 & 0.9172 & 37.2 & 354.9 & 50.9 & 197.0 &  86.10 & 19.36 & 0.10 & $-$0.40 & 0.10 & 0.72 \\
SP 191849 & 225.3542 & 0.7727 & 37.9 & 311.9 & 46.1 & 236.4 &  48.90 & 19.59 & 0.17 & $-$0.34 & 0.15 & 0.77 \\
SP 192758 & 227.2607 & 0.9143 & 45.2 &  41.0 & 22.6 & 235.1 &  67.25 & 19.52 & 0.07 & $-$0.39 & 0.10 & 0.48 \\
SP 196227 & 225.3654 & 0.7734 & 40.5 & 307.3 & 50.6 & 194.3 &  73.32 & 19.75 & 0.27 & $-$0.40 & 0.14 & 1.00 \\
          & 232.2728 & 0.9637 & 92.1 &  67.2 & 46.0 & 188.8 & 113.73 & 21.21 & 0.83 & $-$0.21 & 0.74 & 2.06 \\
SP 204220 & 232.3003 & 0.9628 & 84.7 &  61.9 & 31.3 & 300.7 & 100.88 & 20.06 & 0.29 &  0.05 & 0.30 & 1.11 \\
SP 205556 & 227.2880 & 0.9153 & 40.6 &  28.5 & 40.8 & 335.4 &  33.94 & 18.83 & 0.14 & $-$0.29 & 0.06 & 0.29 \\
SP 208793 & 226.3150 & 0.8515 & 32.5 & 355.7 & 18.6 & 292.6 &  28.71 & 19.19 & 0.18 & $-$0.27 & 0.12 & 0.43 \\
SP 210121 & 226.3278 & 0.8521 & 33.1 & 347.4 & 35.1 & 321.1 &  14.82 & 17.70 & 0.65 &  0.12 & 0.46 & 1.08 \\
SP 214174 & 225.2662 & 0.7680 & 27.8 &   8.2 & 14.2 & 151.9 &  40.09 & 19.80 & 0.16 & $-$0.36 & 0.15 & 0.59 \\
SP 214308 & 225.2807 & 0.7688 & 27.7 & 357.2 & 14.6 & 168.2 &  42.20 & 19.82 & 0.17 & $-$0.38 & 0.12 & 0.56 \\
          & 227.3342 & 0.9168 & 37.0 &   3.0 & 19.3 & 217.4 &  53.89 & 19.45 & 0.10 & $-$0.42 & 0.05 & 0.40 \\
SP 214930 & 225.3337 & 0.7716 & 33.6 & 322.0 & 57.8 & 344.9 &  28.96 & 19.05 & 0.21 & $-$0.26 & 0.12 & 0.56 \\
SP 215512 & 225.3237 & 0.7710 & 31.8 & 327.6 & 37.2 & 344.3 &  10.83 & 18.17 & 0.61 &  0.02 & 0.46 & 1.07 \\
          & 232.3124 & 0.9624 & 81.6 &  59.4 & 37.9 & 339.9 &  76.95 & 19.78 & 0.19 & $-$0.01 & 0.20 & 0.90 \\
SP 215657 & 225.3751 & 0.7740 & 42.9 & 303.6 & 24.6 & 231.5 &  41.10 & 19.78 & 0.16 & $-$0.35 & 0.12 & 0.66 \\
          & 229.3659 & 0.9913 & 46.8 &  16.7 & 25.1 & 231.9 &  68.40 & 19.12 & 0.02 & $-$0.41 & 0.05 & 0.44 \\
SP 216009 & 226.2802 & 0.8500 & 33.6 &  18.3 & 13.0 & 159.1 &  44.31 & 19.50 & 0.18 & $-$0.33 & 0.12 & 0.42 \\
          & 232.2881 & 0.9632 & 87.8 &  64.3 & 12.1 & 187.9 &  94.47 & 20.64 & 0.49 &  0.06 & 0.42 & 1.34 \\
SP 217505 & 225.2919 & 0.7693 & 28.1 & 348.9 & 27.3 & 173.0 &  55.43 & 19.98 & 0.14 & $-$0.39 & 0.13 & 0.66 \\
          & 232.4427 & 0.9590 & 56.2 &  22.0 & 45.0 & 216.2 & 100.15 & 19.48 & 0.02 & $-$0.37 & 0.03 & 0.60 \\
SP 219639 & 227.3194 & 0.9163 & 37.5 &  11.6 & 16.4 & 353.6 &  22.46 & 18.80 & 0.19 & $-$0.29 & 0.11 & 0.36 \\
SP 220172 & 225.3144 & 0.7705 & 30.4 & 333.3 & 22.8 &   7.8 &  16.93 & 18.44 & 0.59 & $-$0.04 & 0.45 & 0.98 \\
\hline
\end{longtable}

\null\newpage
\begin{longtable}{lccrrrrrrrrr}
	\caption{Str{\"o}mgren surface magnitudes of the observed sky patches.}
	\label{tab:uvbylong} \\
\hline
Sky patch & $D$ & lunar & $z$ & $\gamma$ & $\zeta$ & $\alpha$ & $\theta$ & $\mu(y)$ & $b-y$ & $v-b$ & $u-v$ \\
label &  & phase & ($\degr$) & ($\degr$) & ($\degr$) & ($\degr$) & ($\degr$) & (mag/$\sq\arcsec$) & (mag) & (mag) & (mag) \\
\hline \endfirsthead

\multicolumn{12}{c}%
{{\tablename\ \thetable{} -- continued from previous page}} \\
\hline
Sky patch & $D$ & lunar & $z$ & $\gamma$ & $\zeta$ & $\alpha$ & $\theta$ & $\mu(y)$ & $b-y$ & $v-b$ & $u-v$ \\
label &  & phase & ($\degr$) & ($\degr$) & ($\degr$) & ($\degr$) & ($\degr$) & (mag/$\sq\arcsec$) & (mag) & (mag) & (mag) \\
\hline \endhead

\hline \multicolumn{12}{r}{{Continued on next page}} \\
\endfoot

\hline
\endlastfoot

SP 955    & 232.4330 & 0.9592 & 57.4 &  25.6 & 33.7 & 288.0 &  67.23 & 19.38 & 0.05 & 0.24 & 0.52 \\
SP 4670   & 227.3767 & 0.9182 & 39.3 & 338.9 & 50.7 &   0.4 &  18.90 & 18.27 & 0.20 & 0.31 & 0.72 \\
SP 5524   & 227.3905 & 0.9186 & 41.0 & 331.9 &  7.5 & 341.3 &  33.69 & 19.04 & 0.16 & 0.26 & 0.58 \\
SP 6815   & 232.3318 & 0.9618 & 76.8 &  55.1 & 45.1 &  23.5 &  41.60 & 19.00 & 0.13 & 0.39 & 0.84 \\
SP 7795   & 229.3870 & 0.9915 & 45.6 &   6.7 & 10.1 & 173.1 &  55.49 & 18.94 & 0.06 & 0.22 & 0.46 \\
SP 9403   & 232.3444 & 0.9615 & 73.9 &  52.1 & 17.9 & 134.6 &  72.37 & 19.75 & 0.11 & 0.28 & 0.66 \\
SP 9404   & 227.4059 & 0.9192 & 43.5 & 324.7 & 12.2 & 178.7 &  53.99 & 19.41 & 0.14 & 0.21 & 0.52 \\
SP 10101  & 232.3556 & 0.9612 & 71.3 &  49.3 & 18.7 & 146.4 &  74.57 & 19.74 & 0.08 & 0.29 & 0.59 \\
SP 10894  & 227.4207 & 0.9197 & 46.3 & 318.4 & 43.3 & 358.6 &  28.18 & 18.61 & 0.16 & 0.29 & 0.67 \\
SP 14788  & 229.4358 & 0.9920 & 47.0 & 343.5 & 17.5 &   1.6 &  30.73 & 18.42 & 0.08 & 0.30 & 0.56 \\
SP 16581  & 229.4499 & 0.9921 & 48.5 & 337.2 & 33.8 & 359.3 &  20.47 & 18.00 & 0.13 & 0.34 & 0.64 \\
SP 18100  & 232.3689 & 0.9608 & 68.4 &  45.8 & 26.3 &  83.9 &  49.17 & 19.29 & 0.09 & 0.29 & 0.63 \\
SP 20086  & 232.3858 & 0.9604 & 65.0 &  41.1 & 55.2 &  34.0 &  11.58 & 17.23 & 0.40 & 0.64 & 1.04 \\
SP 21166  & 232.4195 & 0.9596 & 59.2 &  30.3 & 42.5 & 173.0 &  94.84 & 19.49 & 0.07 & 0.21 & 0.51 \\
SP 23521  & 229.4222 & 0.9919 & 46.0 & 349.8 & 44.9 & 171.7 &  90.87 & 18.94 & 0.07 & 0.20 & 0.46 \\
SP 24249  & 229.6012 & 0.9938 & 81.5 & 292.0 & 31.2 & 239.8 &  63.93 & 19.30 & 0.12 & 0.40 & 0.79 \\
SP 25169  & 229.5961 & 0.9937 & 80.1 & 293.1 & 29.2 & 231.5 &  67.74 & 19.33 & 0.12 & 0.35 & 0.78 \\
SP 27471  & 229.4974 & 0.9926 & 56.2 & 318.9 & 14.3 & 159.9 &  69.60 & 19.16 & 0.06 & 0.19 & 0.51 \\
          & 233.5038 & 0.9115 & 54.7 &  11.0 & 13.2 & 177.9 &  67.66 & 19.79 & 0.12 & 0.22 & 0.54 \\
SP 28482  & 229.5261 & 0.9929 & 62.3 & 310.0 & 55.9 & 359.8 &  42.81 & 18.25 & 0.06 & 0.29 & 0.67 \\
SP 29138  & 232.4053 & 0.9599 & 61.5 &  35.0 & 53.2 & 176.0 & 105.08 & 19.28 & 0.05 & 0.23 & 0.55 \\
SP 34179  & 229.4683 & 0.9923 & 51.0 & 329.6 & 44.2 &  49.3 &  56.77 & 18.62 & 0.07 & 0.24 & 0.54 \\
SP 34677  & 233.5679 & 0.9096 & 55.4 & 345.3 &  6.3 & 258.8 &  55.22 & 19.61 & 0.16 & 0.19 & 0.55 \\
SP 35042  & 233.5160 & 0.9112 & 54.2 &   6.1 & 21.0 &  35.0 &  36.78 & 19.26 & 0.09 & 0.29 & 0.58 \\
SP 36340  & 233.5242 & 0.9109 & 54.0 &   2.8 & 37.4 &  19.5 &  20.36 & 18.73 & 0.13 & 0.33 & 0.69 \\
SP 48150  & 229.4029 & 0.9917 & 45.4 & 359.0 & 59.2 & 124.7 &  89.88 & 18.59 & 0.02 & 0.21 & 0.56 \\
          & 233.5938 & 0.9088 & 58.0 & 335.5 & 11.4 & 166.4 &  69.18 & 19.75 & 0.13 & 0.21 & 0.50 \\
SP 48467  & 229.5099 & 0.9927 & 58.7 & 314.9 & 43.5 & 163.3 &  98.05 & 19.00 & 0.02 & 0.22 & 0.50 \\
          & 379.2655 & 0.9975 & 68.5 &  84.6 & 40.4 & 191.1 &  83.76 & 18.74 & 0.00 & 0.24 & 0.62 \\
SP 49260  & 229.5411 & 0.9931 & 65.9 & 305.9 & 25.9 & 133.2 &  91.57 & 19.29 & 0.05 & 0.19 & 0.51 \\
SP 49559  & 233.5421 & 0.9104 & 54.1 & 355.5 & 22.5 & 131.1 &  71.33 & 19.79 & 0.12 & 0.24 & 0.51 \\
SP 49850  & 229.4848 & 0.9925 & 53.8 & 323.3 & 39.7 & 121.4 &  91.29 & 19.02 & 0.07 & 0.18 & 0.47 \\
          & 379.2459 & 0.9976 & 74.2 &  88.3 & 15.1 & 216.9 &  83.91 & 19.04 & 0.01 & 0.28 & 0.67 \\
SP 51102  & 229.5822 & 0.9935 & 76.4 & 296.0 & 55.5 &  18.5 &  76.16 & 18.85 & 0.07 & 0.31 & 0.75 \\
SP 54197  & 229.5678 & 0.9934 & 72.6 & 299.3 & 28.5 &  55.7 &  86.55 & 19.31 & 0.06 & 0.23 & 0.59 \\
SP 66446  & 233.5761 & 0.9093 & 56.1 & 342.1 & 48.7 &  43.2 &  47.99 & 19.07 & 0.11 & 0.27 & 0.64 \\
SP 79601  & 233.5503 & 0.9101 & 54.3 & 352.2 & 46.2 & 119.9 &  87.40 & 19.53 & 0.14 & 0.24 & 0.53 \\
SP 81035  & 379.4001 & 0.9962 & 31.8 &  43.9 & 27.1 & 227.7 &  58.88 & 18.58 & $-$0.01 & 0.20 & 0.46 \\
SP 81077  & 379.3049 & 0.9971 & 56.9 &  76.4 & 14.9 & 167.3 &  58.36 & 18.62 & 0.00 & 0.26 & 0.56 \\
SP 82121  & 229.5539 & 0.9932 & 69.0 & 302.6 & 50.3 & 125.0 & 119.26 & 18.82 & 0.02 & 0.25 & 0.63 \\
SP 91427  & 379.3243 & 0.9969 & 51.3 &  71.7 & 57.6 &  17.1 &  44.20 & 17.67 & 0.05 & 0.34 & 0.68 \\
SP 92136  & 379.2761 & 0.9974 & 65.3 &  82.5 & 28.2 & 106.5 &  40.49 & 18.16 & 0.06 & 0.34 & 0.69 \\
SP 93431  & 379.4065 & 0.9961 & 30.5 &  40.5 & 39.9 & 343.5 &  33.07 & 17.92 & 0.03 & 0.26 & 0.57 \\
SP 97991  & 379.2970 & 0.9972 & 59.2 &  78.2 & 44.4 &  57.1 &  22.04 & 17.47 & 0.08 & 0.40 & 0.73 \\
SP 105283 & 379.3483 & 0.9967 & 44.6 &  64.9 & 26.2 & 123.5 &  36.84 & 18.00 & 0.06 & 0.31 & 0.60 \\
SP 106295 & 379.3308 & 0.9968 & 49.5 &  70.0 & 55.1 &  48.3 &  17.97 & 16.81 & 0.18 & 0.47 & 0.82 \\
SP 108927 & 379.3815 & 0.9964 & 36.0 &  52.8 & 47.1 & 173.4 &  70.68 & 18.49 & 0.00 & 0.21 & 0.48 \\
SP 125310 & 379.5550 & 0.9947 & 38.7 & 300.9 & 27.4 & 334.9 &  21.36 & 17.80 & 0.05 & 0.31 & 0.61 \\
SP 127294 & 379.3561 & 0.9966 & 42.5 &  62.4 & 50.2 & 121.7 &  42.54 & 17.80 & 0.04 & 0.32 & 0.61 \\
SP 128726 & 379.5172 & 0.9951 & 30.2 & 318.7 & 14.7 & 158.5 &  44.26 & 18.36 & 0.05 & 0.23 & 0.50 \\
SP 134305 & 379.5616 & 0.9946 & 40.5 & 298.4 & 44.8 & 358.2 &  39.67 & 17.94 & 0.07 & 0.27 & 0.57 \\
SP 147559 & 379.5239 & 0.9950 & 31.5 & 315.0 & 29.5 & 137.9 &  61.07 & 18.52 & 0.07 & 0.21 & 0.47 \\
SP 156623 & 379.6461 & 0.9936 & 64.1 & 276.1 & 13.1 & 175.9 &  67.13 & 18.88 & 0.06 & 0.23 & 0.56 \\
SP 159384 & 379.5853 & 0.9943 & 46.8 & 290.8 & 24.5 & 128.6 &  70.37 & 18.66 & 0.07 & 0.23 & 0.48 \\
SP 161572 & 379.6117 & 0.9940 & 54.2 & 283.8 & 42.5 &  30.2 &  73.93 & 18.69 & 0.02 & 0.19 & 0.48 \\
SP 165016 & 379.6282 & 0.9938 & 58.9 & 280.0 & 18.2 &  69.8 &  74.96 & 18.93 & 0.03 & 0.20 & 0.49 \\
SP 171141 & 379.5919 & 0.9943 & 48.6 & 288.9 & 33.1 & 125.9 &  80.62 & 18.84 & 0.01 & 0.18 & 0.45 \\
SP 183570 & 227.2798 & 0.9150 & 41.8 &  32.5 & 43.1 & 281.3 &  67.70 & 19.14 & 0.16 & 0.22 & 0.55 \\
SP 189247 & 226.3048 & 0.8511 & 32.4 &   2.4 & 36.5 & 238.3 &  59.98 & 19.39 & 0.22 & 0.24 & 0.54 \\
SP 189502 & 227.3106 & 0.9160 & 38.1 &  16.6 & 38.5 & 237.7 &  70.93 & 19.29 & 0.17 & 0.21 & 0.52 \\
SP 189723 & 227.3540 & 0.9174 & 37.4 & 351.4 & 51.4 & 197.2 &  86.14 & 19.16 & 0.25 & 0.19 & 0.57 \\
SP 192758 & 227.2665 & 0.9145 & 44.1 &  38.5 & 24.0 & 236.3 &  67.30 & 19.41 & 0.14 & 0.20 & 0.53 \\
SP 196227 & 232.2795 & 0.9635 & 89.9 &  65.9 & 46.3 & 189.5 & 113.44 & 20.50 & 0.67 & 0.53 & 0.75 \\
SP 204220 & 232.3052 & 0.9626 & 83.4 &  60.9 & 32.6 & 298.5 & 100.95 & 19.86 & 0.20 & 0.40 & 0.88 \\
SP 205556 & 227.2942 & 0.9155 & 39.8 &  25.4 & 41.7 & 332.3 &  33.99 & 18.73 & 0.14 & 0.28 & 0.61 \\
SP 208793 & 226.3202 & 0.8517 & 32.7 & 352.3 & 20.1 & 290.0 &  28.75 & 19.07 & 0.19 & 0.30 & 0.59 \\
SP 210121 & 226.3335 & 0.8523 & 33.5 & 343.9 & 36.2 & 318.1 &  14.87 & 17.69 & 0.43 & 0.57 & 0.93 \\
SP 214174 & 225.2718 & 0.7683 & 27.7 &   4.0 & 13.5 & 157.0 &  40.11 & 19.65 & 0.25 & 0.25 & 0.52 \\
SP 214308 & 227.3399 & 0.9170 & 37.0 & 359.7 & 20.4 & 220.0 &  53.93 & 19.32 & 0.15 & 0.23 & 0.50 \\
SP 215512 & 232.3179 & 0.9622 & 80.2 &  58.2 & 38.5 & 336.8 &  77.01 & 19.61 & 0.18 & 0.32 & 0.80 \\
SP 215657 & 229.3720 & 0.9914 & 46.3 &  13.8 & 26.5 & 233.1 &  68.44 & 19.03 & 0.04 & 0.19 & 0.48 \\
SP 216009 & 226.2865 & 0.8503 & 33.1 &  14.4 & 12.4 & 166.0 &  44.34 & 19.39 & 0.18 & 0.28 & 0.56 \\
          & 232.2936 & 0.9630 & 86.4 &  63.2 & 12.4 & 194.5 &  94.58 & 20.30 & 0.36 & 0.44 & 0.83 \\
SP 219639 & 227.3251 & 0.9165 & 37.3 &   8.3 & 16.7 & 346.8 &  22.50 & 18.73 & 0.18 & 0.30 & 0.61 \\
\hline
\end{longtable}

\end{center}
%%%%%%%%%%%%%%%%%%%%%%%%%%%%%%%%%%%%%%%%%%%%%%%%%%

% Don't change these lines
\bsp	% typesetting comment
\label{lastpage}
\end{document}